%% file: stirLC.tex
\documentclass[twocolumn,times,trackchanges]{aastex63}

\usepackage{graphicx}
\graphicspath{{./}}
\usepackage{amsfonts,amsmath,amssymb}
\usepackage{xspace}
\usepackage{bm,booktabs}
\usepackage{longtable,topcapt,rotating }
\usepackage{multirow}

\newcommand{\Msun}{\ensuremath{\mathrm{M}_\odot}\xspace}

\received{February 3, 2021}
\accepted{\today}
\submitjournal{ApJ}

\shorttitle{Type IIP CCSN Light Curves and Progenitors} 
\shortauthors{Barker et al.}

\begin{document}

\title{Connecting the Light Curves of Type IIP Supernovae to the Properties of their Progenitors}

\author[0000-0002-8825-0893]{Brandon L.~Barker}
\altaffiliation{NSF Graduate Research Fellow}
\affiliation{Department of Physics and Astronomy, Michigan State University, East Lansing, MI 48824, USA}
\affiliation{Department of Computational Mathematics, Science, and Engineering, Michigan State University, East Lansing, MI 48824, USA}

\author[0000-0002-1751-7474]{Chelsea E.~Harris}
\affiliation{Department of Physics and Astronomy, Michigan State University, East Lansing, MI 48824, USA}

\author[0000-0001-9440-6017]{MacKenzie L.~Warren}
\affiliation{AI Research Group, 5x5 Technologies, St Petersburg, FL 33701 USA}

\author[0000-0002-8228-796X]{Evan P.~O'Connor}
\affiliation{The Oskar Klein Centre, Department of Astronomy, Stockholm University, AlbaNova, SE-106 91 Stockholm, Sweden}

\author[0000-0002-5080-5996]{Sean M.~Couch}
\affiliation{Department of Physics and Astronomy, Michigan State University, East Lansing, MI 48824, USA}
\affiliation{Joint Institute for Nuclear Astrophysics-Center for the Evolution of the Elements, Michigan State University, East Lansing, MI 48824, USA}
\affiliation{Department of Computational Mathematics, Science, and Engineering, Michigan State University, East Lansing, MI 48824, USA}
\affiliation{National Superconducting Cyclotron Laboratory, Michigan State University, East Lansing, MI 48824, USA}

\correspondingauthor{Brandon L.~Barker}
\email{barker49@msu.edu}

\begin{abstract}
  Observations of core-collapse supernovae (CCSNe) reveal a wealth of information about the dynamics of the supernova ejecta and its composition but very little direct information about the progenitor. 
  Constraining properties of the progenitor and the explosion requires coupling the observations with a theoretical model of the explosion. 
  Here, we begin with the CCSN simulations of \citet{couch:2020}, which use a non-parametric treatment of the neutrino transport while also accounting for turbulence and convection. 
  In this work we use the SuperNova Explosion Code to evolve the CCSN hydrodynamics to later times and compute bolometric light curves. 
  Focusing on SNe IIP, we then (1) directly compare the theoretical STIR explosions to observations and (2) assess how properties of the \textit{progenitor's core} can be estimated \textit{from optical photometry in the plateau phase alone}. 
  First, the distribution of plateau luminosities (L$_{50}$) and ejecta velocities achieved by our simulations is similar to the observed distributions. 
  Second, we fit our models to the light curves and velocity evolution of some well-observed SNe. 
  Third, we recover well-known correlations, as well as the difficulty of connecting any one SN property to zero-age main sequence mass. 
  Finally, we show that there is a usable, linear correlation between iron core mass and L$_{50}$ such that optical photometry alone of SNe IIP can give us insights into the cores of massive stars. 
  Illustrating this by application to a few SNe, we find iron core masses of 1.3-1.5 solar masses with typical errors of ~0.05 solar masses. 
  Data are publicly available online (\url{https://doi.org/10.5281/zenodo.6631964}).
\end{abstract}


\section{Introduction}

Core-collapse supernovae (CCSNe) are the explosive deaths that result from the ends of stellar evolution for massive stars with zero-age main sequence (ZAMS) masses $M_{\mathrm{ZAMS}} \gtrsim 8M_{\odot}$.
The current understanding suggests that some fraction of possible progenitors will successfully produce CCSNe while others will fail and produce a black hole (BH) \citep{oconnor:2011,lovegrove:2013,ertl:2016,sukhbold:2016,adams:2017a,sukhbold:2018,couch:2020}.
The details of the explosion mechanism have been the subject of decades of work with current work favoring, for most progenitors, the delayed neutrino-driven mechanism \citep{bethe:1985}.
For an in-depth review of the CCSN explosion mechanism and related problems, see recent reviews \citep[e.g.,][]{bethe:1990, janka:2007, janka:2012a, janka:2016, burrows:2013, hix:2014, muller:2016, couch:2017, pejcha:2020}.

CCSNe are detectable by three primary messengers -- EM waves, neutrinos, and gravitational waves (GWs).
Neutrino and GW signals have the very desirable property that they are emitted directly from the core of the star at the time of collapse and may reveal information about the structure there \citep[e.g.,][]{pajkos:2019, pajkos:2020, warren:2020, sotani:2020a}, unlike photons which are emitted from the photosphere in the outer layers of the supernova ejecta until the remnant phase.
However, to date there has been only one detection of neutrinos from a supernova \citep[][SN1987A]{arnett:1989}. 
With modern neutrino detectors, only CCSNe occurring within our galaxy may be detectable \citep{scholberg:2012}. 
Similarly, there have been no confirmed detections of GW emission from a CCSNe. 
The current suite of detectors (aLIGO, Virgo, and KAGRA) can only detect GWs from a CCSNe if it occurs within a distance of $\leq 100$ kpc \citep{abbott:2016a}. 
It is the case, however unfortunate, that the overwhelming majority CCSNe will only be observed in EM signals. 

The focus of this paper is connecting EM signals to progenitor properties for SNe IIP.
These events have been shown to originate from red supergiant progenitors \citep{van-dyk:2003, smartt:2009, van-dyk:2019}.
Despite being the most common type of CCSNe, their diversity of observable features -- such as light curve morphologies -- is still not fully understood \citep[e.g., ][]{anderson:2014, valenti:2016}. 
The connection between SNe IIP and IIL supernovae, for example, still remains an open question -- whether IIL's are the limit of IIP's as the H envelope is depleted or a separate class \citep{barbon:1979, blinnikov:1993, faran:2014a, morozova:2015}.

Understanding the connection between SNe IIP light curves and stellar progenitors has a new urgency. 
Coming next-generation telescopes such as the Vera C. Rubin Observatory and its primary optical photometry survey, The Rubin Observatory Legacy Survey of Space and Time (LSST) \citep{ivezi:2019}, will allow for extremely deep imaging of the entire sky every couple of nights. 
The LSST will allow for statistical studies of populations of CCSNe of an unprecedented scale \citep[for recent statistical studies see, e.g.,][]{anderson:2014, sanders:2015, gutierrez:2017, gutierrez:2017a}. 

Ultimately, properly characterizing the diversity in SNe II supernova light curve morphology will require the union of observation and theory.
On the theory side, this comprises realistic stellar evolution models including the core collapse, following the resulting explosion with robust physics, and calculating EM light curves (as well as neutrino and GW signals). 
The gold standard is full three-dimensional (3D), self-consistent simulations.
Core-collapse supernovae and their progenitors are truly 3D in nature and the key to understanding the diversity of light curve morphology lies in faithfully modeling these asphericities \citep{wongwathanarat:2013,wongwathanarat:2015,dessart:2019a,stockinger:2020,sandoval:2021}.
3D simulations are, however, computationally expensive to perform and, as such, are limited in number and the range of parameter space that they cover.
Spherically symmetric (1D) simulations remain necessary for understanding the CCSNe explosion mechanism and their observables by surveying landscapes of possible CCSNe.
Great progress has been made in the last few years regarding 1D CCSN simulations \citep{ebinger:2017, sukhbold:2016, couch:2020}, allowing for successful explosions in 1D using neutrino-driven explosions across wide ranges of progenitor masses.
These 1D simulations allow for large parameter studies performing potentially thousands of simulations spanning ranges of progenitor masses, equations of state, and metallicities, for example.

Light curve calculations are the final, crucial piece of the theoretical process of understanding these explosions.
Commonly, calculations of synthetic bolometric light curves of CCSNe invoke a thermal bomb or piston-driven model, where energy is artificially injected into a thin region above a user-specified mass cut within the progenitor \citep[see, e.g., ][]{bersten:2011, morozova:2015, ricks:2019}.
In these models, the explosion energy is a user-set parameter instead of being determined by the structure of the progenitor and explosion physics.
The calculations cannot determine whether a given progenitor will result in a successful supernova or fail to revive its stalled shock and collapse to a black hole.
The explodability has been shown to have non-trivial behavior across a large range of ZAMS mass progenitors \citep{sukhbold:2016, ebinger:2017, sukhbold:2018, couch:2020} and cannot be captured with more simplified models.
The clear next step is the coupling of high fidelity CCSN simulations with bolometric light curve calculations.

Light curves contain information about their progenitor and the explosion -- properties such as the composition of the ejecta, mass or radius of the progenitor, or explosion energy may be inferred \citep{litvinova:1985, popov:1993, kasen:2009a, sukhbold:2016}.
The process of inferring progenitor and explosion properties from light curves has been shown to be highly degenerate \citep{goldberg:2019, dessart:2019} with many combinations of properties being capable of producing a given light curve.
Of particular interest, however, is the early time light curve dominated by radiation streaming form the shock heated outer envelope.
This early time behavior can be compared to shock cooling models to put constraints on the stellar radius \citep{nakar:2010, tolstov:2013, shussman:2016, kozyreva:2020a}.
Recently \citet{morozova:2016c, rubin:2017} explored the effectiveness and temporal limitations of these models and these methods have been widely used for constraining the progenitor pre-explosion radius \citep[e.g.,][]{rabinak:2011, gall:2015, gonzalezgaitan:2015, sapir:2016, soumagnac:2020, vallely:2021}.
These early time observations may help to break the degeneracies between progenitor and explosion properties \citep{goldberg:2020b}.

In this work, we calculate the bolometric light curves of the recent 1D simulations done with the \texttt{FLASH}\footnote{\url{https://flash.rochester.edu/site/}} \citep{fryxell:2000, dubey:2009} code using the new Supernova Turbulence In Reduced-dimensionality (STIR) model \citep{couch:2020}.
This 1D convection scheme has the benefit of being more consistent with some properties of full physics 3D CCSN simulations -- such as explosion energies and landscapes -- while leaving the neutrino physics unaltered.
Like any 1D method, it remains a simplification of the full picture and is not without its shortcomings \citep[e.g.,][]{mueller:2019}.
Similar 1D schemes have also been used to study Rayleigh-Taylor instabilities in supernova remnants \citep{duffell:2016}.
The initial conditions of these models are set by the 1D stellar evolution models of \citet{sukhbold:2016}, which make up a suite of 200 solar metallicity, non-rotating massive stars between 9 and 120 \Msun. 
We couple the final state of the STIR simulations with the SuperNova Explosion Code (SNEC) \citep{morozova:2015}, which follows the explosion through the rest of the star and through the plateau and nebular phases of the light curves.
We will demonstrate that using a more sophisticated 1D explosion model to determine a distribution of explosion energies consistent with 3D simulations imparts non-trivial features to observables and thus properties inferred from them, highlighting the importance of the explosion model used.
With this set of light curves, we make available a new set of theoretical predictions to compare directly with observations.
Furthermore, we investigate direct correlations between progenitor properties and light curve properties.
We recover known correlations, and we quantify the dependence of SNe IIP luminosity on the progenitor iron core mass at time of collapse -- thus providing a way of obtaining core properties from EM signals without the need for the much rarer neutrino and GW signals. 

This paper is laid out as follows: in Section~\ref{sec:methods} we discuss the various progenitors, codes, and statistical methods that are used in this study. 
Section~\ref{sec:Results} presents our results: \ref{sec:landscapes} presents observable properties of our light curves and their trends across ZAMS mass, \ref{sec:observations} presents preliminary comparisons to observations of SNe IIP, \ref{sec:correlations_results} shows correlations found between light curve and progenitor properties.
In Section~\ref{sec:Conclusions}, we summarize our results and briefly discuss comparison to other theoretical light-curve calculations and prospects for future work. 

\section{Methods}
\label{sec:methods}

For this work, we begin with massive stellar progenitors evolved up to the point of core collapse in \citet{sukhbold:2016} evolved using KEPLER.
The core collapse and following explosion or collapse to BH are simulated using the \texttt{FLASH} simulation framework \citep{fryxell:2000, dubey:2009} with the Supernova Turbulence In Reduced-dimensionality (STIR) model \citep{couch:2020}, the details of which are discussed in Section~\ref{sec:FLASH}. 
The output of the STIR simulations are mapped into the SuperNova Explosion Code (SNEC) \citep{morozova:2015, morozova:2016c, morozova:2018b} to generate bolometric light curves as discussed in Section~\ref{sec:SNEC}. 
In Section~\ref{sec:Correlations} we present the methods used to analyze statistical relationships between properties of the progenitor and observables.

\subsection{Progenitors}
\label{sec:progenitors}

We begin with the 200 non-rotating, solar metallicity models of \citet{sukhbold:2016}. 
These models cover a range of ZAMS masses from 9 -- 120 M$_{\odot}$ and were created with the KEPLER code assuming no magnetic fields or rotation and single star evolution.
Progenitors with ZAMS masses above 31\Msun experienced significant mass loss during their lifetimes and did not explode as SNe IIP and this is the upper limit on our mass range \citep[see][for details on their stellar evolution]{sukhbold:2016}.
The more massive Type I SNe progenitors are too few in number to perform a meaningful statistical analysis and we defer their analysis for future work.
This leaves 136 progenitors producing SNe IIP supernovae used in this work.

\begin{figure}
  \centering
  \includegraphics[width = 0.45\textwidth]{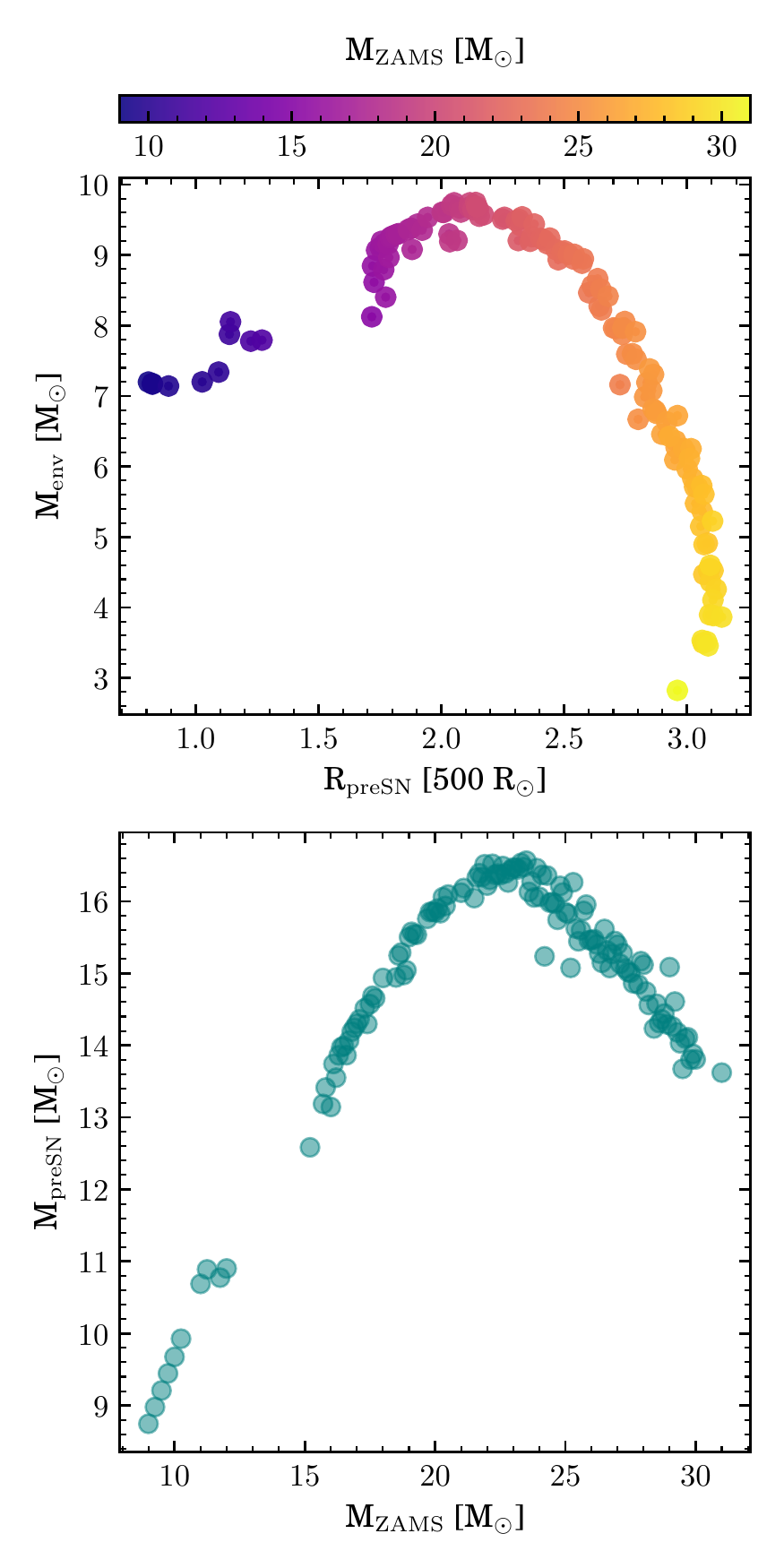}
  \caption{Properties of the progenitors of \citet{sukhbold:2016}. (top) Mass of the H-rich envelope ($M_{\mathrm{env}}$) as a function of pre-supernova radius ($R_{\mathrm{preSN}}$).
    (bottom) Final stellar mass ($M_{\mathrm{preSN}}$), after mass loss, as a function of ZAMS mass.}
  \label{fig:progenitors}
\end{figure}

These progenitors span a wide range of possible CCSNe progenitor properties.
Figure~\ref{fig:progenitors} shows the mass of the H-rich envelope as a function of pre-supernova radius (top) and the stellar pre-supernova mass as a function of ZAMS mass (bottom).
Here we show only models that successfully exploded in \citet{couch:2020} and are included in this work. 
Gaps in this figure -- such as that from about 12-15\Msun -- represent models which failed to explode and are not included in this work.
These progenitors become mass-loss dominated around 23\Msun, as seen in the bottom panel of Figure~\ref{fig:progenitors}.
This complicates correlations between quantities of interest and tends to cause them to deviate from monotonicity.
This is key to investigating observable trends in light curves across a wide range of progenitors, as we demonstrate later.

The progenitors in \citet{sukhbold:2016} were further investigated in \citet{sukhbold:2018} using a set of high resolution stellar evolutionary models.
They showed that the features of these progenitors -- notably the compactness landscape -- was not numerical in nature and was present in their high resolution models.
Similarly, other, recent works have found similar trends in the presupernova mass and compactness \citep[e.g.,][using \texttt{MESA} \citep{paxton:2011, paxton:2013, paxton:2015, paxton:2018, paxton:2019}]{laplace:2021}.
We note that while general trends may be reproduced, details such as the apparent `chaos' seen in \citet{sukhbold:2016, sukhbold:2018} are sensitive to implementation details of stellar evolution and may not appear in other studies \citep{chieffi:2020}.
Using a different set of progenitors with a different compactness curve would likely affect the explosion landscape but isn't expected to affect the results presented here.

A key result of systematic 1D studies of CCSNe are the so-called ``islands of explodabilty'' \citep{sukhbold:2016}. 
The final result of stellar core collapse - a successful or failed explosion - is not a monotonic function of ZAMS mass. 
Instead, the explodability of the progenitor is sensitive to the core structure at the time of collapse. 
While the placement of these islands is sensitive to the explosion model and the progenitors used, it is a feature that has now been seen amongst many groups \citep{oconnor:2011, perego:2015, sukhbold:2016, ertl:2016, ebinger:2018, couch:2020}.
However, studies using exclusively thermal bomb driven explosions uninformed by neutrino driven calculations cannot reproduce the explosion/implosion fate of a progenitor and are insensitive to this feature.
Any systematic study of light curves from populations of SNe must capture this complex behavior.

\subsection{\texttt{FLASH}}
\label{sec:FLASH}

The CCSN simulations were conducted in \citet{couch:2020} using the \texttt{FLASH} code framework with the STIR turbulence-aided explosion model. 
This model is a new method for artificially driving explosions in 1D CCSN simulations.. 
Turbulence is key in simulating successful, realistic explosions, as turbulence may constitute 50$\%$ or more of the total pressure behind the shock \citep{murphy:2013, couch:2015a} and turbulent dissipation is important for post-shock heating \citep{mabanta:2018}. 
The combined impact of these effects is to aid the explosion. 
The inclusion of turbulent effects allows for successful explosions in 1D simulations while reproducing the results seen in 3D simulations from various groups \citep{couch:2020} without the need for parametrized neutrino physics.

STIR models turbulence using the Reynolds-averaged Euler equations with mixing length theory as a closure. 
This model has one primary scalable parameter, the mixing length parameter $\alpha_{\Lambda}$, inherited from mixing length theory which scales the strength of convection.
The mixing length parameter has been tuned to fit STIR simulations to full 3D simulations run with \texttt{FLASH} and reproduces 3D results seen in \texttt{FLASH} and other codes, noting particularly good agreement with the 3D works of \citep{burrows:2020}.
We use the fiducial value found in \citet{couch:2020} for the mixing length parameter, $\alpha_{\Lambda} = 1.25$.
STIR also includes four additional diffusion parameters that control the convective mixing of internal energy, turbulent kinetic energy, composition, and neutrinos. 
As in \citet{couch:2020}, all four of these diffusion coefficients are set to 1/6, a value consistent with comparison to fully 3D simulation of convection in massive stars \citep{muller:2016}.
We note that the convective dynamics are insensitive to the choice of diffusion coefficients and, thus, impacts on the explosion are negligible \citep{muller:2016, boccioli:2021}.
\texttt{FLASH} with the STIR model has the desireable benefit that there is no need to tune the model to match a specific observation.
Instead, its one primary parameter is tuned to be consistent with multi-physics 3D CCSN simulations, reducing the possibility of inserting biases into the results.

STIR includes neutrino transport using a state-of-the-art two moment method with an analytic ``M1'' closure \citep{shibata:2011, cardall:2013, oconnor:2015, oconnor:2018}.
We simulate three neutrino flavors: $\nu_{e}$, $\bar{\nu}_{e}$ and $\nu_x$, where $\nu_x$ combines the $\mu$ -- $\tau$ neutrino and antineutrino flavors.
M1 transport requires no tuning and has no free parameters (up to the choice of a closure for the high-order radiation moments), allowing for truly physics-driven explosions.
The STIR simulations use the now commonly adopted, empirically-motivated ``SFHo'' equation of state for dense nuclear matter \citep{steiner:2013} which is able to replicate observed neutron star masses.

At the end of the STIR simulations, the explosion energies for all but the highest-mass progenitors have asymptoted. 
It is commonplace in CCSNe work to define the explosion energy as the sum total energy, from all sources, of material that has both positive total energy and positive velocity \citep[e.g., ][]{bruenn:2016} at the end of the simulation.
This is zero during the stalled shock phase, when all of the material is still gravitationally bound, and becomes positive if/when the shock begins to move outward again due to neutrino heating and other effects.
This energy, once it has reached its asymptotic value, represents the energy that is injected into the rest of the star to drive the explosion and unbind the stellar material.
When discussing the combined STIR + SNEC simulations, this is the explosion energy that we will reference.
It is important to note that this energy is different than the energy that would be used in hydrodynamical modelling (e.g., thermal bomb explosions).
In the thermal bomb regime, a user set energy is deposited at $t=0$ over defined temporal and spatial extent, and assumes that the energy of the shock comes directly from the core-bounce which is inconsistent with the physical picture of CCSNe.
In the case of high fidelity simulations, a large amount of material has already been gravitationally unbound by the shock when the explosion energy is measured.
A thermal bomb model with the ``same'' energy injected into the inner zones would, by the time the same amount of material is unbound, be less energetic by exactly the binding energy of the material.
Care should be taken when comparing energetics from these two approaches.
While the physics of these two explosion methods are inconsistent with each other, the thermal bomb energetics can be made consistent with neutrino-driven energetics by correcting the bomb energy by the binding energy of the material between the shock and the PNS surface (this material is already unbound when the explosion energy is calculated as above, but in thermal bomb or piston-driven explosions it is not).
Without this correction, a thermal bomb model using energetics from neutrino-driven simulations will have less energy available for the explosion, impacting observables.

\begin{figure}
  \centering
  \includegraphics[width = 0.45\textwidth]{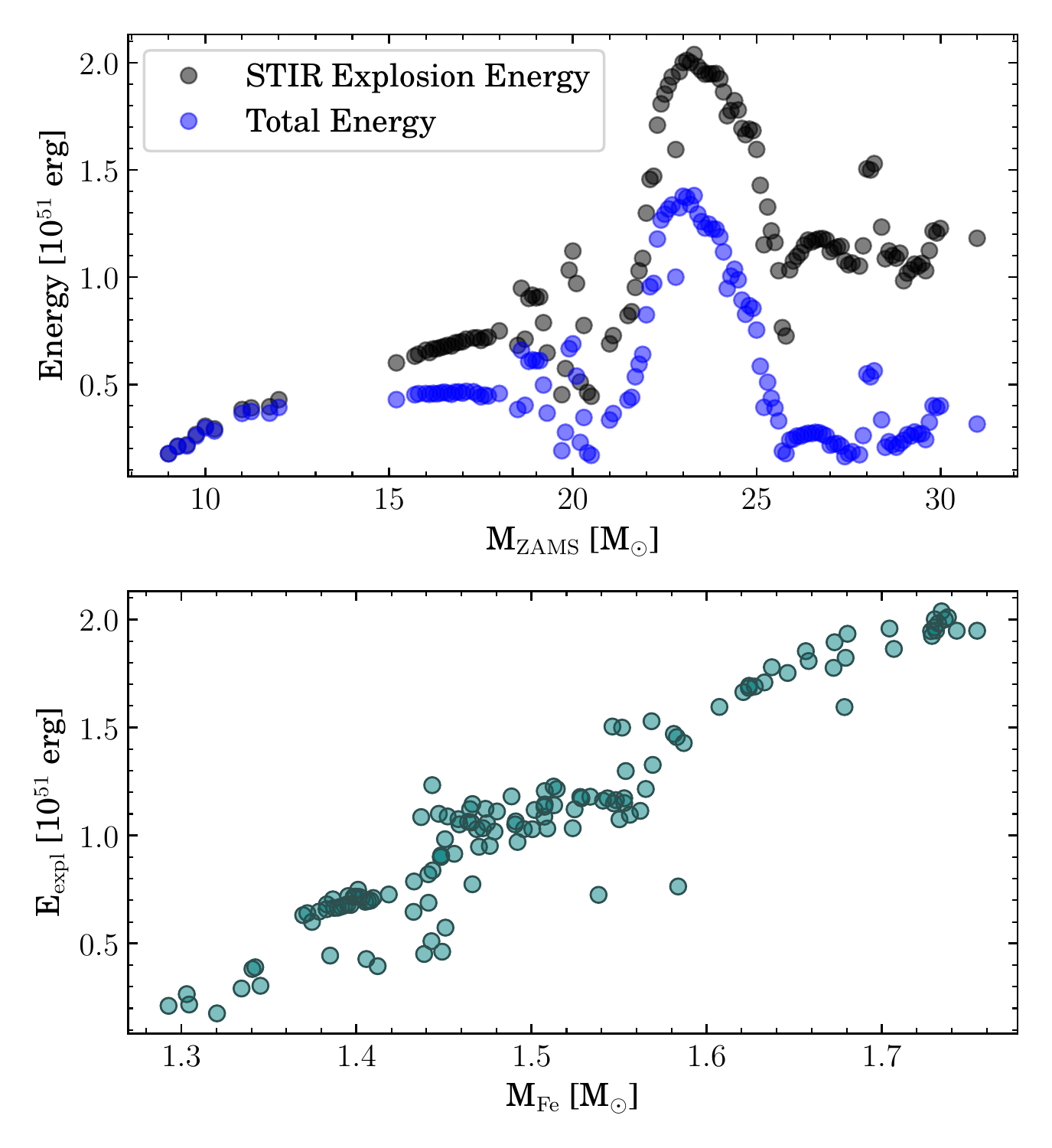}
  \caption{Top: Explosion energies realized in the STIR CCSN simulations of \citet{couch:2020} (black) and the final energy after removing the overburden energy of the progenitor (blue). 
  Bottom: STIR explosion energy as a function of the progenitor's iron core mass.}
  \label{fig:e_expls}
\end{figure}

Figure~\ref{fig:e_expls} shows the explosion energies obtained with STIR (black) alongside the explosion energy with the progenitor's overburden energy removed (blue).
The progenitor's overburden energy is the (negative) total energy above the shock that the explosion must overcome to unbind the star \citep{bruenn:2016}.
The total energy, which we compute as the total energy on the computational domain after the explosion has set in is closer to what will characterize the ejecta.
Gaps in mass, such as from about 12\Msun to 15\Msun, indicate regions where progenitors failed to successfully launch an explosion in STIR.
The bottom panel shows the explosion energy as a function of the iron core mass.
These explosion energies are set largely by the structure of the cores of their progenitors -- effects which can only be seen by employing neutrino driven explosions \citep[for recent examples of the impacts of core structures on explosions and observable signatures, see][]{warren:2020, burrows:2020}.
The emerging picture from high fidelity simulations is that there is no simple relationship between explosion energy and ZAMS mass, instead requiring multi-physics simulations to determine robustly \citep{sukhbold:2016,ebinger:2017, ebinger:2018, sukhbold:2018, couch:2020, burrows:2020, ertl:2020}.
The explosion energy is more closely related to the pre-supernova mass and properties of the core, such as the compactness parameter or the mass of the iron core.

\subsection{SNEC}
\label{sec:SNEC}

We simulate light curves for all of the models that successfully produced explosions in \citet{couch:2020} (see their Figure~6, middle row).
This is all but about 50 of the original 200 progenitors.
This limits our study to light curves obtained from progenitors that actually explode, allowing us to explore solely relationships that come from physically-driven explosions.
At the end of the STIR simulations, the final states are mapped into the SuperNova Explosion Code (SNEC) \citep{morozova:2015}. 
SNEC is a spherically symmetric, Lagrangian, equilibrium flux-limited diffusion radiation-hydrodynamics code and is publicly available\footnote{\url{http://stellarcollapse.org/SNEC}}.
Unlike STIR, SNEC does not include any form of general relativistic gravity, neutrino transport, or dense matter EoS, which are all important for modeling the explosion but not necessarily for computing the light curve.
Instead, it follows the basic physics needed for predicting bolometric supernova light curves.
SNEC includes Lagrangian Newtonian hydrodynamics with artificial viscosity following the formulation in \citet{mezzacappa:1993a} 
and a stellar equation of state following \citet{paczynski:1983} that includes contributions from radiation, ions, and electrons with approximate electron degeneracy. 
This is used in tandem with a Saha ionization solver that can follow ionization of any number of present elements.
At high temperatures SNEC uses OPAL Type II opacities \citep{iglesias:1996} suitable for solar metallicity.
These opacities are supplemented by those of \citet{ferguson:2005} at low temperatures.

1D modeling cannot properly capture the mixing at compositional interfaces due to Rayleigh-Taylor and Richtmyer-Meshkov instabilities, for example.
Without mixing, sharp compositional gradients appear that produce features in light curves that are not observed in nature \citep{utrobin:2007}.
In these mixing processes, shock propagation outwards can cause light elements to mix inwards and heavy elements to mix outwards \citep{wongwathanarat:2015}.
Of particular importance is the mixing of radioactive $^{56}$Ni, whose mixing extent affects the light curve properties \citep{morozova:2015}.
SNEC applies boxcar smoothing that smooths out compositional profiles, simulating mixing and avoiding unphysical light curve bumps.
We use the fiducial parameters of \citet{morozova:2015} for our boxcar smoothing.

In the present work we follow the ionization of $^{1}$H, $^{3}$He, and $^{4}$He, similarly to \citet{morozova:2015}.
H and He make up the majority of the energy contributions from recombination relevant for producing bolometric SNe IIP light curves.
Our STIR simulations do not currently track detailed compositional information in their output. 
When mapping into SNEC, we fill the composition in the STIR part of the domain to be pure $^{4}$He.
This has no noticeable effect on the light curves in this study (see Appendix~\ref{app:composition}).
Figure~\ref{fig:composition} shows mass fractions of $^{1}$H (light blue), $^{4}$He (dark blue), $^{12}$C (gold), $^{16}$O (red), and $^{56}$Ni (black, dot-dashed line).
The solid lines show the unmixed profiles that are input to SNEC.
Notably, the gray region shows the STIR domain where the composition, prior to mixing, is set to pure $^{4}$He.
The dashed lines show the compsotion after boxcar smoothing is applied.
The bottom panel shows the radial density profile in the STIR domain (solid line) and in SNEC after mapping (dashed line).

\begin{figure}
  \centering
  \includegraphics[width = 0.45\textwidth]{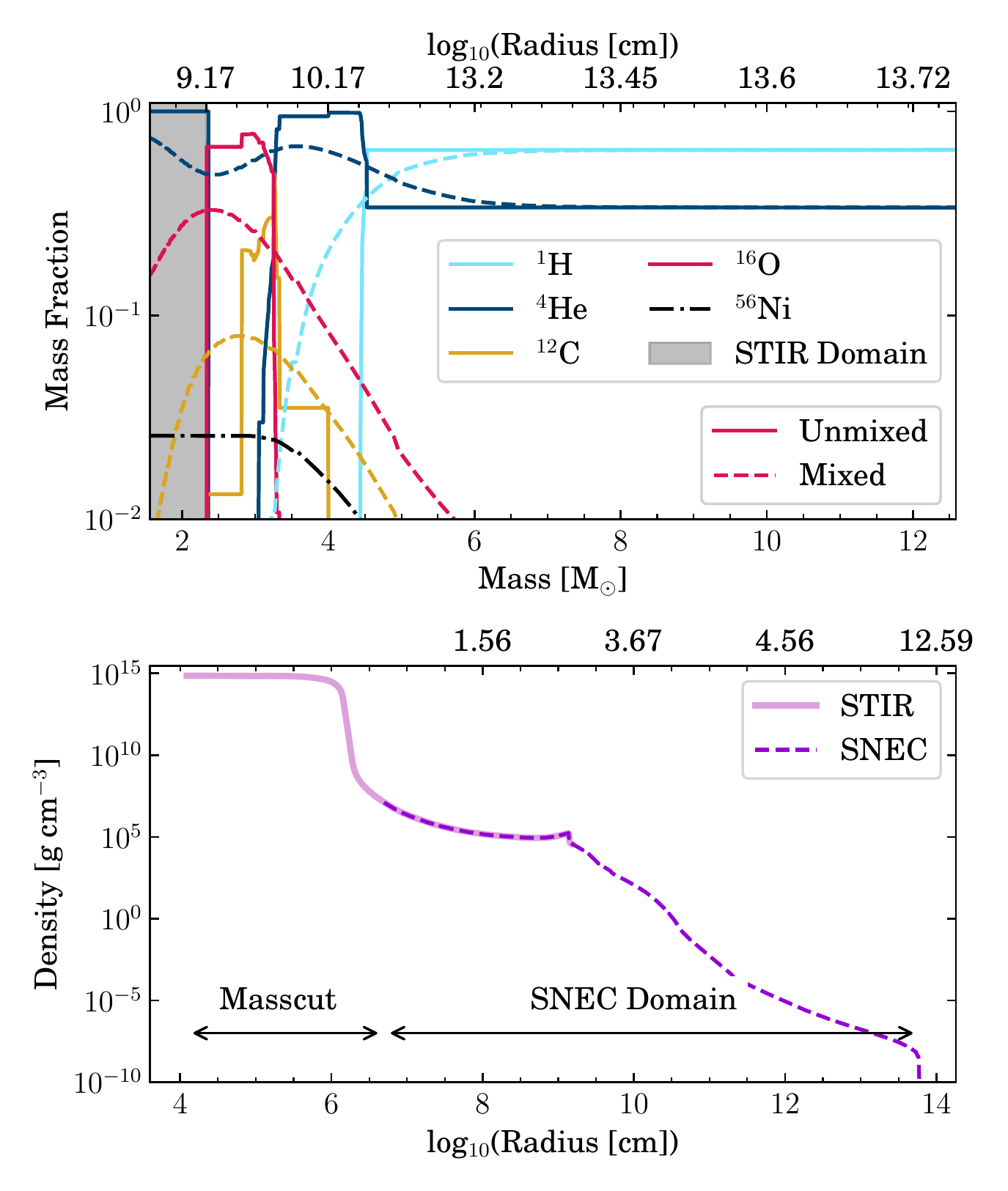}
  \caption{Top: Mass fraction of $^{1}$H (light blue), $^{4}$He (dark blue), $^{12}$C (gold), $^{16}$O (red), and $^{56}$Ni (black, dot-dashed line).
  Solid lines show the unmixed profiles, dashed lines show the profiles after boxcar smoothing is applied. The gray shaded region represents the STIR domain, which is originally set to pure $^{4}$He prior to smoothing.
  Bottom: Radial density profile for the STIR domain (solid line) and SNEC mapping (dashed line). }
  \label{fig:composition}
\end{figure}

The final, critical ingredient for powering a SNe light curve is radioactive heating from the $^{56}$Ni $\rightarrow$ $^{56}$Co $\rightarrow$ $^{56}$Fe decay chain.
Radioactive $^{56}$Ni is produced in explosive nuclear burning during the first epochs of the explosion in the inner parts of the star.
Hydrodynamic instabilities mix the $^{56}$Ni outward.
Gamma-rays and positrons emitted from the decay process diffuse outward and provide an additional source of energy.
Capturing this is crucial as, after the end of the plateau phase, the light curve is powered entirely by this radioactive decay.
SNEC follows the radiative transfer of gamma-rays from the $^{56}$Ni and $^{56}$Co decays using the gray transfer approximation \citep{swartz:1995} and the resulting energy release is coupled to the hydrodynamics independently from the rest of the radiation. 

Currently, neither our STIR models used here nor the public version of SNEC include a nuclear reaction network. 
To alleviate this issue, SNEC allows for a user specified amount of $^{56}$Ni to be input by hand throughout a specified mass coordinate. 
\citet{sukhbold:2016} simulate the explosions of these progenitors including a large nuclear reaction network, and we use the $^{56}$Ni yields as a function of explosion energy from their work (see their Table 4, Figure 17) to estimate a mass of $^{56}$Ni from that relationship to be distributed by SNEC.
For all but the lightest progenitors, they find around 0.07\Msun of $^{56}$Ni. 
We disperse the $^{56}$Ni up to about 75$\%$ of the way through the He shell -- avoiding mixing into the H envelope.
This provides control amongst the progenitors. 
As the mixing extent must be set by hand, any further treatment would require a large parameter study.
In recent high fidelity models, mixing of radioactive $^{56}$Ni into the H-rich envelope is realized \citep{utrobin:2015a, utrobin:2017, stockinger:2020, utrobin:2021a, sandoval:2021} and is expected to occur in at least some of our models. 
In \citet{morozova:2015}, they showed that variations in these distributions had little effect on the light curve, especially on the plateau (see their Figure~6). 
\citet{goldberg:2019} show slight variations in the light curve as it falls off the plateau depending on the extent of mixing (see their Figure~10).
\citet{kozyreva:2019} explore the effects of mixing prescriptions for $^{56}$Ni, such as uniform or boxcar, on light curves, showing differences on the plateau between these methods.
The lack of a reaction network consistently incorporated into the calculations forms a weakness of the current work, despite being based on nucleosynthetic calculations and tuned to our explosion energies.
However, the main results of this work (see Section~\ref{sec:correlations_results}) use quantities measured on the plateau where they are less sensitive to reasonable variations in $^{56}$Ni mass and distribution.
Future work will include nucleosynthesis calculations with the STIR input models to properly seed the SNEC calculations.

Typically, high fidelity CCSN simulations do not simulate the entire star -- instead focusing on the inner 15,000 km or so necessary for launching the explosion.
We must stitch the STIR simulation data, with the explosions developed on the grid, onto the progenitor pre-explosion profile outside the STIR boundary (15,000 km) in order to simulate the full star.
Below the shock, mass profiles are taken from STIR.
Above the shock mass coordinate, mass profiles are taken from the progenitor profiles.
These smooth, combined STIR -- pre-explosion progenitor profiles are used as the inputs to SNEC as detailed in \citep{morozova:2015}.
One advantage of using the STIR models as the initial conditions to SNEC is that the high fidelity equation of state and neutrino transport yield a physically realistic remnant mass to motivate the mass cut -- an amount of material not included in light curve simulations that should be close to the remnant mass. 
We place a mass cut outside the PNS at the point where the total energy becomes positive -- removing both the PNS and a small amount of still gravitationally bound material above it (of order 0.0001\Msun).
For all of the simulations we use 1000 cells in the SNEC domain using a geometric grid, as in \citet{morozova:2015}, that places higher resolution in the core around the shock and at the outer domain to resolve the photosphere.
Our grid is slightly modified from that of \citet{morozova:2015} to place added resolution in the core over the already existing explosion.
Simulations were run until 300 days when possible to adequately sample both the plateau and the tail for all events.

To simulate CCSNe directly from progenitors, SNEC has the ability to artificially drive an explosion with a piston or thermal bomb.
One of the primary qualities of our method is to eliminate the need for this and thus eliminate user input explosion energies, which can take any range or distribution, replacing them with physically motivated energetics.
However, for some of the more massive progenitors in this study, the explosion energies were still increasing by the time the shock reached the outer boundary. 
Eventually, energy generation from neutrino heating and other sources will slow as the shock expands and the explosions energies will asymptote.
Since our computational domain is limited to 15,000 km, some progenitors do not reach their ``true'' explosion energies.
In order to fully capture the energy of the explosion in STIR, we integrate the neutrino heating in the gain region at the end of the \texttt{FLASH} simulations to estimate the asymptotic explosion energy and add the difference -- at most about 0.3$\times 10^{51}$ erg -- as a thermal bomb over the shocked region.
These additions are most necessary in the region of high energy between about 21\Msun and 25\Msun where the final energies were still readily increasing.
This energy is what is displayed in Figure~\ref{fig:e_expls}.

The light curves presented in this work represent those 136 progenitors (of the suite of 200) that
both successfully launch an explosion (Section~\ref{sec:FLASH}) and have light curves that would
be identified as a SNe IIP, which we find is simply a mass cut of $M_\mathrm{ZAMS} \leq 31$\Msun.

\subsection{Correlations}
\label{sec:Correlations}

We are interested in uncovering correlations between observable properties of the explosion and properties of the progenitors.
The size and fidelity of the sample allows us to address these connections necessary to understand light curve diversity.
Our robust treatment of the explosion physics combined with large sample of progenitors makes us uniquely situated to address correlations in a novel way. 
We proceed similarly to \citet{warren:2020}, wherein the correlations between observed neutrino and GW signals with progenitor properties were addressed.

We measure correlations with the Spearman's rank correlation coefficient.
The Spearman correlation coefficient measures any monotonic relationships between variables, in contrast to the Pearson coefficient which measures only linear correlations.
It is important that we are able to access non-linear relationships that are seen in the data.
The combined effect of a wide range of stellar progenitors with mass loss effects and non-linear, non-monotonic explosion energetics over the range of progenitors produces robust and realistic -- but not necessarily linear -- relationships.

The Spearman coefficient is obtained by first ranking the data by replacing the values by their indices after sorting.
For example, the data (1.5\Msun, 1.4\Msun, 1.6\Msun) would transform to (2, 1, 3). 
Then, the Spearman rank correlation coefficient is obtained by computing the Pearson correlation of the transformed data, calculated by 

\begin{equation}
\rho = \frac{ \sum_{i}\left( x_{i} - \bar{x}\right) \left( y_{i} - \bar{y}\right) }{\sqrt{\sum_{i}\left( x_{i} - \bar{x}\right)^2} \sqrt{\sum_{i}\left( y_{i} - \bar{y}\right)^2} }
\end{equation}  

\noindent for ranked variables $x$ and $y$ with $\bar{x}$ and $\bar{y}$ being the mean values.
This process of first ranking the data is what allows the Spearman process to produce a more robust correlation metric.
We note that the above equation is the same as that for the Pearson correlation coefficient and, when used on non-ranked data, will produce the Pearson correlation coefficient.

A value of +1(-1) represents an exact monotonic correlation (anticorrelation) and a value of 0 indicates no monotonic relationship.
We consider values $\lvert \rho \rvert \gtrsim 0.5$ to indicate strong statistical correlation, values $0.3 \lesssim \lvert \rho \rvert \lesssim 0.5$ to be moderate correlation, and $\lvert \rho \rvert \lesssim 0.3$ to be a weak correlation, as is standard practice.
Correlation coefficients were calculated using Python's \textit{scipy.stats.spearmanr} package.

We strived to limit observables considered to those reasonably detectable with current facilities -- mostly photometric and early time (meaning, in this context, on the plateau but not requiring observations within days of explosion) features. 
Ultimately plateau duration, plateau luminosity, and ejecta velocity -- all at early times --  proved to be the most useful and accessible.
We explored numerous properties of the progenitors for correlations with observables -- shell masses, density structures, core compactness, and envelope mass to name a few.
Most of these parameters had weak relationships with observable properties.
Ultimately, we settled on the mass of the iron core as the most meaningful and useful progenitor property, as we will see in the next section.

\subsection{Light curve fitting}
\label{sec:fitting}

A common method for estimating CCSN progenitor properties is to construct a grid of models with varying masses, explosion energies, and $^{56}$Ni masses and distributions and select the progenitor from that grid that best fits an observed light curve \citep[see][for recent examples]{morozova:2018b,martinez:2019,martinez:2020}.
We accomplish this by finding the progenitor which minimizes the average relative error $\varepsilon$ of a quantity $f$
\begin{equation}
\varepsilon(f) = \frac{1}{N}\sum_{t^*=t_1}^{t_N} \frac{\lvert f_{t^*} - f^{*}_{t^*}\rvert }{f^{*}_{t^*}}
\end{equation}
\noindent where $f^{*}_{t^*}$ is the observed quantity at time $t^*$, $f_{t^*}$ is the synthetic quantity at the same time, and $N$ is the number of observational data points.
We compare the synthetic and observed data only at the times where observational data is available, using the closest synthetic data to the observational data, which is always within 0.02 days with the output frequency used with SNEC. 
That is, we do not interpolate between observational data points.
We not not consider uncertainties in the explosion epoch in the current work.
We seek models that match both observed bolometric luminosity and velocity evolution, i.e., we seek a model minimizing the combined error metric $\varepsilon(L_{50})+\varepsilon({v_{\mathrm{Fe}}})$.

Other approaches have been used, such as (historically) simply fittng by eye, $\chi^2$ minimization \citep{morozova:2018b}, and Markov chain Monte Carlo methods \citep{martinez:2020}.
We implemented several minimization approaches and found that the above method worked best for the current work.
This is discussed more in Section~\ref{sec:fit_lcs}.

\section{Results}
\label{sec:Results}

We consider the properties of the bolometric light curves followed through the end of the plateaus and into the radioactive tails and the ejecta velocities for models with ZAMS masses 9M$_{\odot}$ $\leq$ M$_{\mathrm{ZAMS}}$ $\leq$ 31M$_{\odot}$ for a total of 136 progenitors. 
In an effort to find relationships with observables that are easily detectable, we consider primarily the photometric and spectroscopic properties in the plateau phase. 
The primary quantities that we consider are the plateau luminosity at day 50 ($L_{50}$), the plateau duration ($t_{p}$), and the ejecta velocity at day 50 ($v_{50}$). 
These quantities are commonly used when inferring explosion properties from observations \citep[e.g., ][]{litvinova:1985, popov:1993, pejcha:2015c} and so their trends from realistic models are of particular interest.
These quantities are easily detectable by current and next generation facilities without the need for late time observations or particularly high cadences, acknowledging that the photosperic velocity will not be as easily observable for most sources. 
This will allow for a relationship to be obtained between these quantities and properties of the core of the progenitor that is both robust and easily detectable with standard measurements. 

\subsection{Landscape properties across ZAMS mass}
\label{sec:landscapes}

Here we present global trends in photometric properties to test the impact of our explosion calculation on light curve features.
As we will see, these properties exhibit non-monatonic features as a function of ZAMS mass and thus introduce degeneracy into attempts to infer progenitor properties from direct comparisons to light curves.

\begin{figure}
  \centering
  \includegraphics[width = 0.45\textwidth]{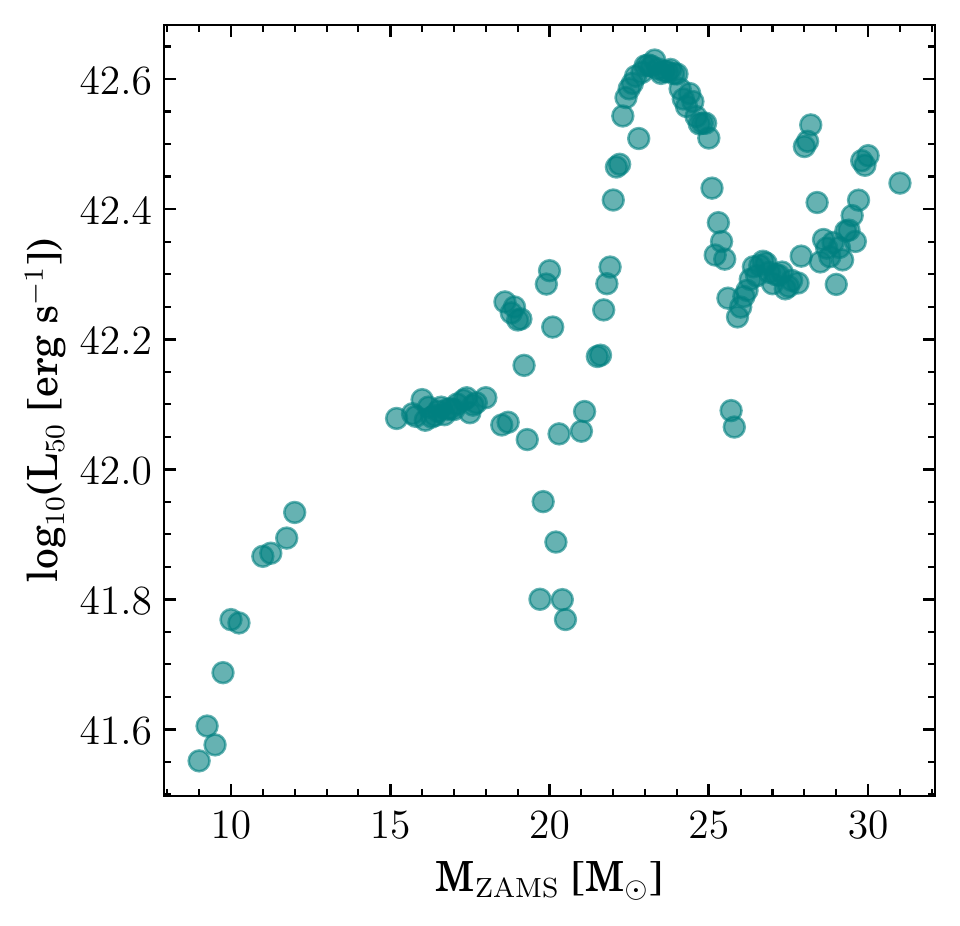}
  \caption{Log of the plateau luminosity at day 50 for the STIR + SNEC models. }
  \label{fig:l50_comp}
\end{figure}

Figure~\ref{fig:l50_comp} shows the bolometric luminosity at day 50 (on the plateau for all progenitors) for all masses. 
The imprint from the distribution of explosion energies is readily seen in the plateau luminosities, with more energetic explosions yielding brighter plateaus.
A consequence of this is the highly degenerate mapping between plateau luminosity and ZAMS mass following the explosion energy distribution (Figure~\ref{fig:e_expls}).

\begin{figure}
  \centering
  \includegraphics[width = 0.45\textwidth]{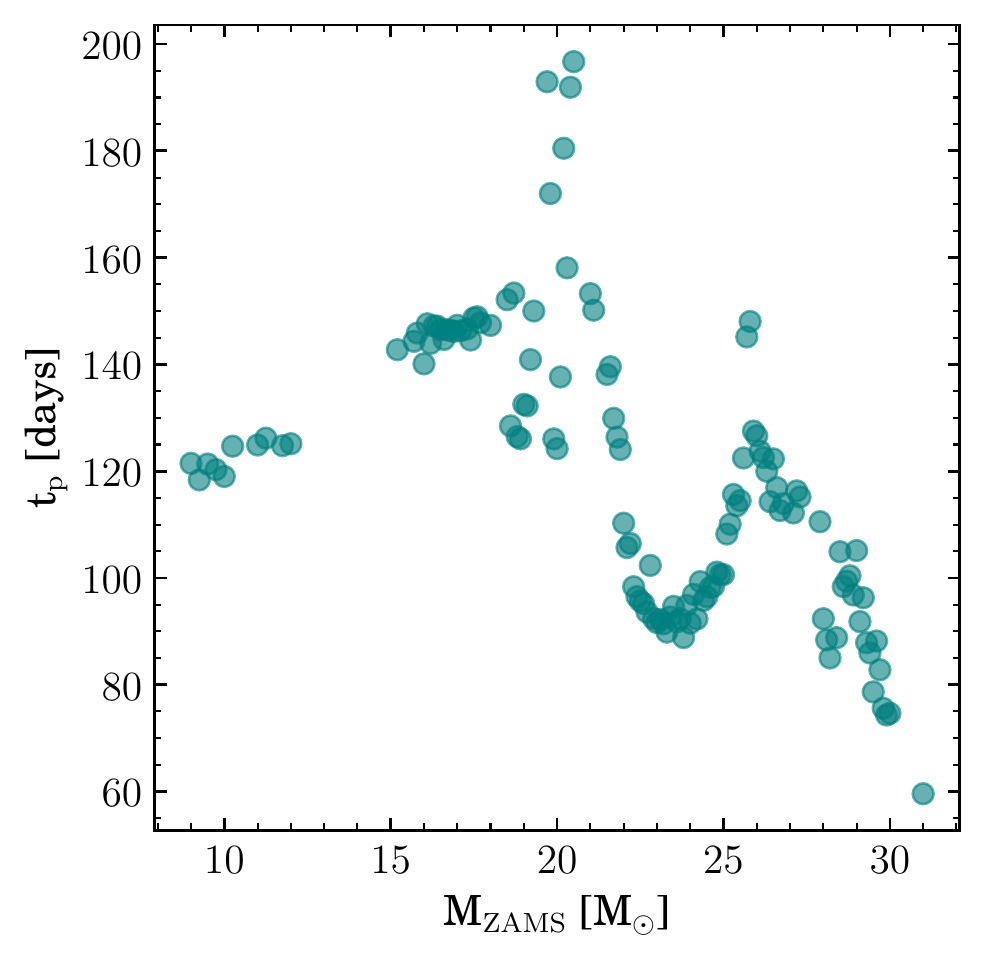}
  \caption{Plateau duration for the STIR + SNEC models. 
  Two progenitors between 27 and 28\Msun have been removed for fair comparison, as some of them did not reach the radioactive tail in the simulation time.}
  \label{fig:tp_comp}
\end{figure}

Figure~\ref{fig:tp_comp} shows the plateau duration for the STIR + SNEC models.
We follow \citet{valenti:2016} and \citet{goldberg:2019} and compute the plateau duration by fitting part of the light curve near the end of the plateau to a combined Fermi-Dirac -- linear function of the form 

\begin{equation}
  \label{eq:plateau}
f(t) = \frac{-a_{0}}{1 + \mathrm{exp}({(t - t_{p})/w_0}) } + (p_{0}\,t) + m_0.
\end{equation}

This avoids biases or inconsistencies that may possibly be introduced by determining the plateau durations by eye for a large sample of light curves.
The physical significance of the various fitting parameters is described in detail in \citet{valenti:2016} and \citet{goldberg:2019}.
Importantly, the parameter $t_{p}$ is taken to be the plateau duration and tends to be placed about halfway through the drop off of the plateau.
Also of interest are $a_0$ and $w_0$ which describe the luminosity drop at the end of the plateau and the width of the drop, respectively.
Fitting was done using Python's \textit{scipy.optimize.curvefit} package starting shortly before the end of the plateau.
For a few of the high mass models between 27 and 28\Msun, timestep restrictions made it difficult to simulate the explosions into the radioactive tails.
Most made it to the end of the plateau and began drop off, but two progenitors were unable to reach the end of the plateau.
For the former case, the fitting is unable to work properly and the plateau duration is set by hand in a way that was consistent with the fitting routine.
For the two progenitors that could not reach the end of the plateau -- 27.4\Msun and 27.5\Msun -- we omit them in comparisons involving the plateau duration.

Clearly, the distribution of the explosion energies imparts a resulting morphology on the plateau durations that cannot be reproduced without energetics informed by neutrino-driven explosions.
We note that many of the plateaus here are quite long, greater than 150 days or so, which is not very common.
These plateaus originate from very massive progenitors, around 20\Msun, which are rare in nature.
Moreover, these models retain quite massive H-rich envelopes (see Figure~\ref{fig:progenitors}) and have reduced explosion energies (see Figure~\ref{fig:e_expls}).
The combination of massive H-rich envelope with reduced explosion energy results in extended plateaus \citep{popov:1993}.
Some uncertainty in the plateau duration remains through the prescription for setting the mass and mixing of radioactive $^{56}$Ni, as it lengthens the plateau slightly \citep{kasen:2009a, morozova:2015, sukhbold:2016, goldberg:2019, kozyreva:2019}. 
These uncertainties, however, should be on the order of days (see, e.g., Figure~13 from \citet{morozova:2015}, Figure~10 from \citet{goldberg:2019}, Figure~4 from \citet{kozyreva:2019}).
It is also somewhat difficult to fairly compare plateau durations to observed works, as many authors present the length of the optically thick phase duration \citep[e.g.,][]{gutierrez:2017a} which may be smaller than our measurement by another 5-10 days or more.
For these reasons, we defer further comparisons to observational data of the plateau durations to future work.

All of this directly impacts the ability to reliably extract progenitor features from light curves.
Without a distribution of explosion energies that is set by a physically realistic explosion model, any sort of arbitrary distribution of light curve properties may be recovered, even with the same diversity of progenitors used.
While STIR is not a perfect or parameter free description of the explosion -- no 1D model ever will be -- it matches well with 3D results and provides a large set of such physically motivated explosion energies for these studies.

\begin{figure}
  \centering
  \includegraphics[width = 0.45\textwidth]{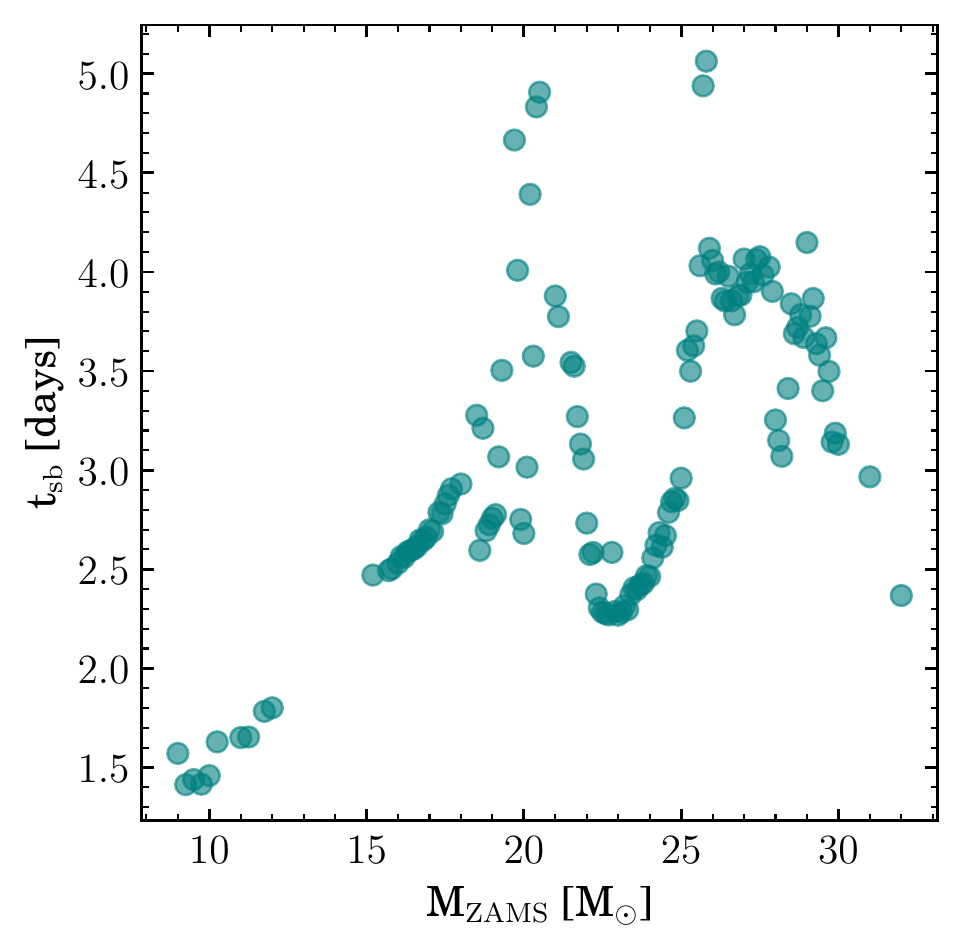}
  \caption{Time for shock breakout for the STIR + SNEC models. }
  \label{fig:tsb_comp}
\end{figure}

Another quantity of interest -- albeit not a directly observable one -- is the time to shock breakout.
Figure~\ref{fig:tsb_comp} shows the time for the shock to breakout from the stellar surface for the STIR + SNEC models.
This is particularly important, as the time to shock breakout sets the on source window for electromagnetic follow-ups of gravitational wave and neutrino events from core-collapse supernovae \citep{abbott:2019c}.
The time to shock breakout is sensitive to the structure of the progenitor and the explosion energy and may be significantly over- or under-estimated if an incorrect explosion energy is used.

With the next galactic CCSNe and prospects for detecting their gravitational wave and neutrino signals, the time to shock breakout becomes a measurable quantity through the difference between GW or neutrino detection time and first light from the SNe.
The SuperNova Early Warning System (SNEWS) \citep{adams:2013, kharusi:2021} will alert observatories to trigger an EM followup after a neutrino detection, and knowing the shock breakout time will be an important factor for the followup study.
Combined with constraints from the GW detection \citep{abbott:2019c} and constraints from other EM observations, the time to shock breakout could help to place additional constraints on the SNe progenitor -- provided that adequate energetics are used.
Similarly, constraints on the shock breakout time after an EM signal may be used to look back at GW and neutrino data, assuming a nearby event.

The previous figures highlight the strong dependence on the distribution of explosion energies used to drive the explosion.
This leads to degeneracies when mapping from observables to ZAMS mass with many progenitors of varying masses being capable of producing a given observation.

\subsection{Comparisons with observations}
\label{sec:observations}

In this section, we compare our light-curves to observations of SNe IIP both through global properties of 
many SNe and fits to the light-curves of individual SNe~IIP that have $M_\mathrm{ZAMS}$ determined through 
pre-explosion imaging data.

\subsubsection{Comparison with a large observational sample}

\begin{figure}
  \centering
  \includegraphics[width = 0.45\textwidth]{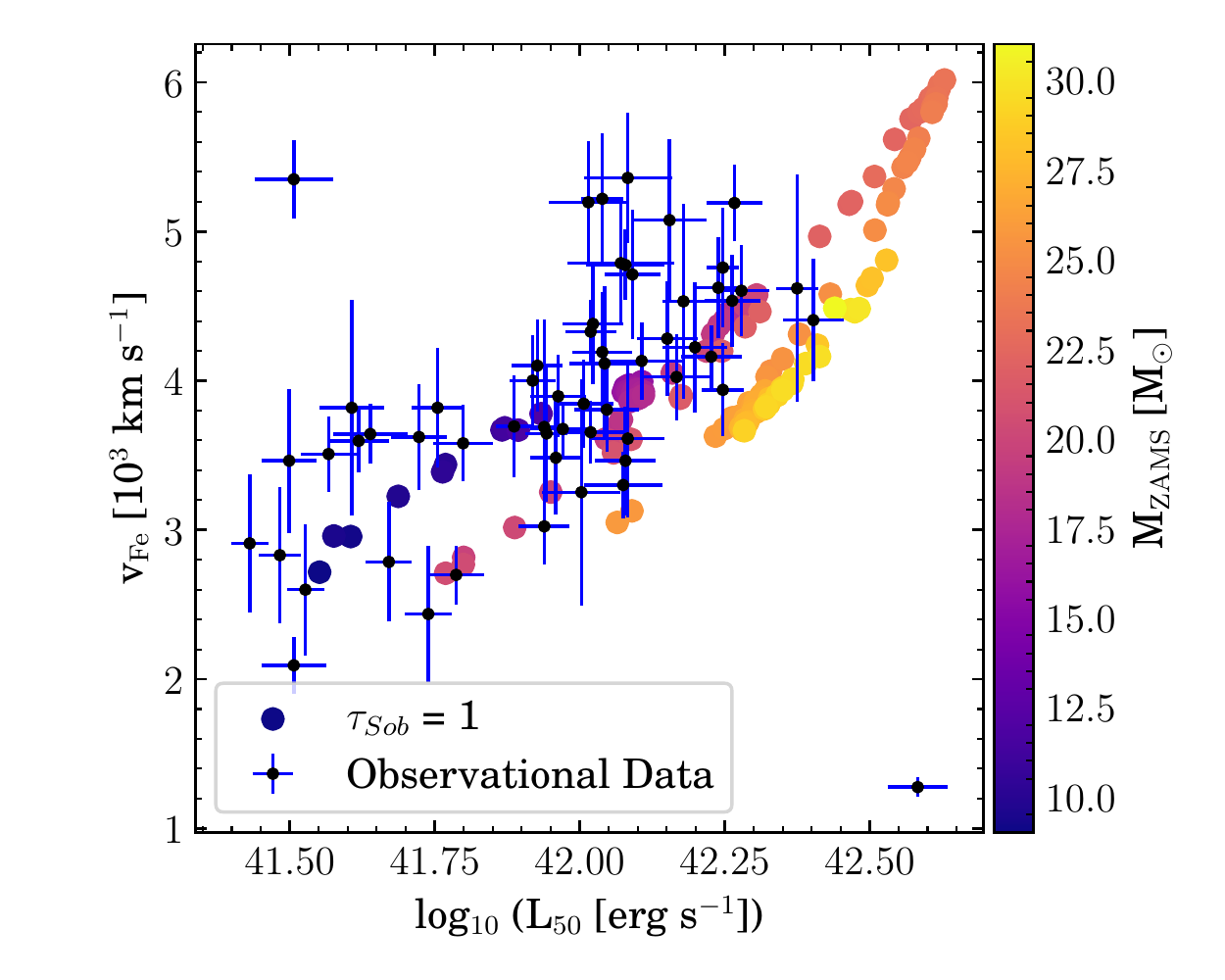}
  \caption{Ejecta velocity at day 50, $v_{50}$, versus the log of the bolometric luminosity of the plateau at day 50, 
  $L_{50}$ for all of the exploding progenitors. Simulated data are colored by the zero-age main sequence mass.
  Points with error bars are observational data from \citet{gutierrez:2017, gutierrez:2017a}.}
  \label{fig:v50_l50}
\end{figure}

Apart from photometric observations, spectroscopic observations may also be used to constrain progenitor properties. 
While we have not computed full synthetic spectra in this work, we can approximate standard line velocities.
Figure~\ref{fig:v50_l50} shows ejecta velocity at day 50 ($v_{50}$) versus plateau luminosity at day 50 ($L_{50}$) for all progenitors that exploded as SNe IIP.
Also plotted are data presented in \citet{gutierrez:2017, gutierrez:2017a}\footnote{Bolometric luminosity data were calculated from $M_{V}$ measurements at day 50 provided by C. Guti\`errez (private communication).}. 
All ejecta velocities here are inferred from the Fe II ($5169\AA$) line. 
In our models, this velocity is calculated in post-processing as the velocity of the ejecta at the point where the Sobolev optical depth ($\tau_{\mathrm{Sob}}$) is unity, with

\begin{equation}
  \tau_{\mathrm{Sob}} = \frac{\pi q_{e}^2}{m_{e} c}\, n_{\mathrm{Fe}} \, \eta_{i} f\, t_{\mathrm{expl}}\, \lambda_{0}
\end{equation}

\noindent where $q_{e}$ and $m_{e}$ are the electron charge and mass, $n_{Fe}$ is the number density of iron atoms, $\eta_{i}$ is the ionization fraction relevant for the transition of interest, $f = 0.023$ is the atomic oscillator strength, $t_{\mathrm{expl}}$ is the time since explosion, and $\lambda_{0}$ is the wavelength associated with the transition.
For material in homologous expansion, this measures the strength of a particular line \citep{mihalas:1978, kasen:2006} and the point where $\tau_{\mathrm{Sob}}=1$ has been shown to match better to observational measurements than the $\tau = 2/3$ electron scattering photosphere \citep{goldberg:2019, paxton:2018}.
To estimate the ionization fraction, we use a table of $\eta_{i}$ as a function of density and temperature that is now publicly available in \texttt{MESA} \citep{paxton:2018}.

We choose to use this metric for the velocity evolution because ultimately we seek to compare with observables.
While the standard $\tau=2/3$ photosphere -- and its velocity -- are simple to compute, they are not simple to observe.
On the other hand, the Fe II $5169\AA$ line is commonly measured.
Therefore, we seek to estimate the location in the ejecta where this line is measured, using the Sobolev approximation that has been readily used in recent works \citep{paxton:2018,goldberg:2019,martinez:2020}.
However, this approach to estimating the iron line velocity is ultimately an approximation and there are physical uncertainties associated with this method.
\citet{paxton:2018} investigated the effects of the choice of the Soboloev optical depth used and found relatively small differences when compared to using the traditional photospheric velocity.
In lieu of full spectral calculations this method provides an estimate of the desired velocity but more work may be needed to robustly compare to observed ejecta velocities.

The sample of luminosities and velocities from our models matches well with the observational sample, but reach higher in luminosity than the observed set. 
These high luminosity events are from some of the higher pre-supernova mass stars around the transition to the mass-loss dominated regime (see Figure~\ref{fig:progenitors}).
These high mass stars are less common than their lower mass companions. 
The highest ZAMS mass stars, those $\gtrapprox$ 23\Msun dominated by mass-loss, dip back down and left in luminosity- and velocity-space, obtaining similar luminosities but slightly lower velocities than lower mass progenitors.
Ultimately, we are able to reproduce observed distributions quite well without having to tune to observations, instead following the explosions from self-consistent simulations.

\subsubsection{Determination of progenitor properties for individual events}
\label{sec:fit_lcs}

It is commonplace to estimate supernova progenitor parameters using a grid hydrodynamical models (i.e., codes similar to SNEC using a thermal bomb) with varying initial masses, thermal bomb energies, and other parameters, and determining the best fitting model \citep[see, e.g.,][]{utrobin:2008, utrobin:2009, pumo:2017, morozova:2018b, martinez:2019, martinez:2020, eldridge:2019}.
We attempt to match our set of explosions with 7 observed bolometric light curves from \citet{martinez:2019, martinez:2020}\footnote{Observational data were provided by L. Martinez (private communication).}. 
Bolometric luminosities are calculated using the bolometric correction method of \citet{bersten:2009}, which requires only BVI photometry to estimate the bolometric correction.

\begin{figure*}
  \centering
  \includegraphics[width = 1.\textwidth]{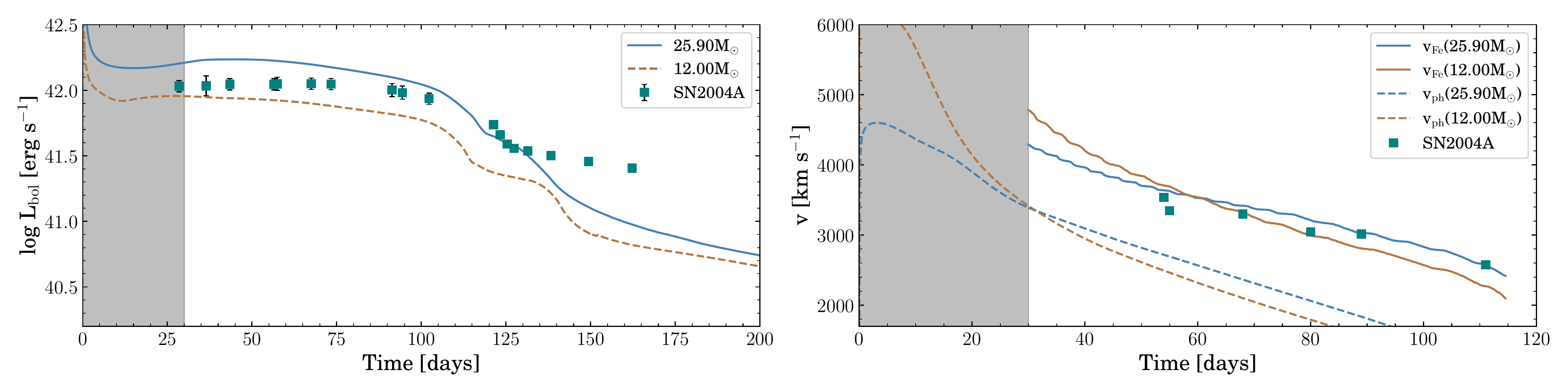} \hfill
  \includegraphics[width = 1.\textwidth]{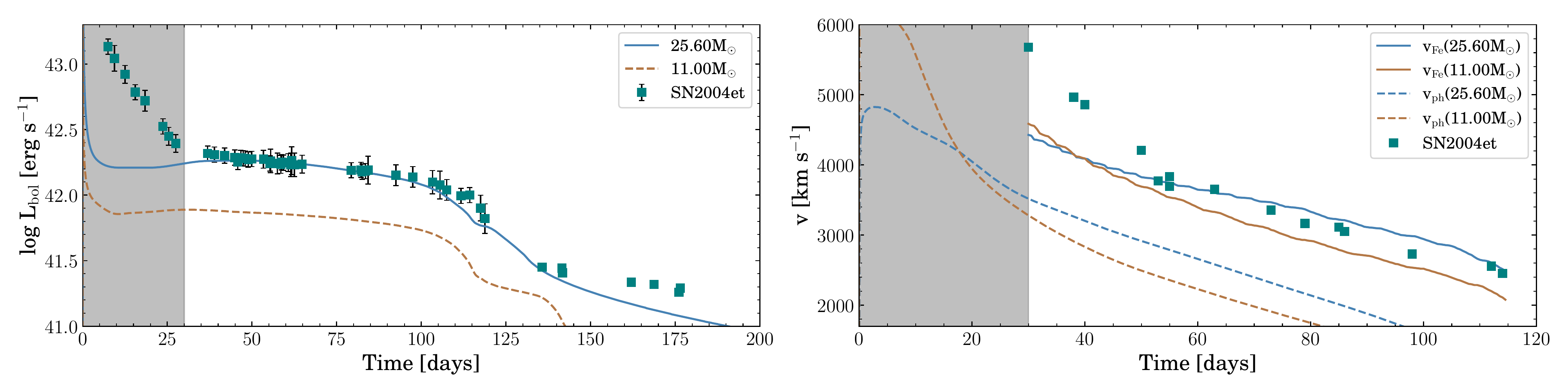} \hfill
  \includegraphics[width = 1.\textwidth]{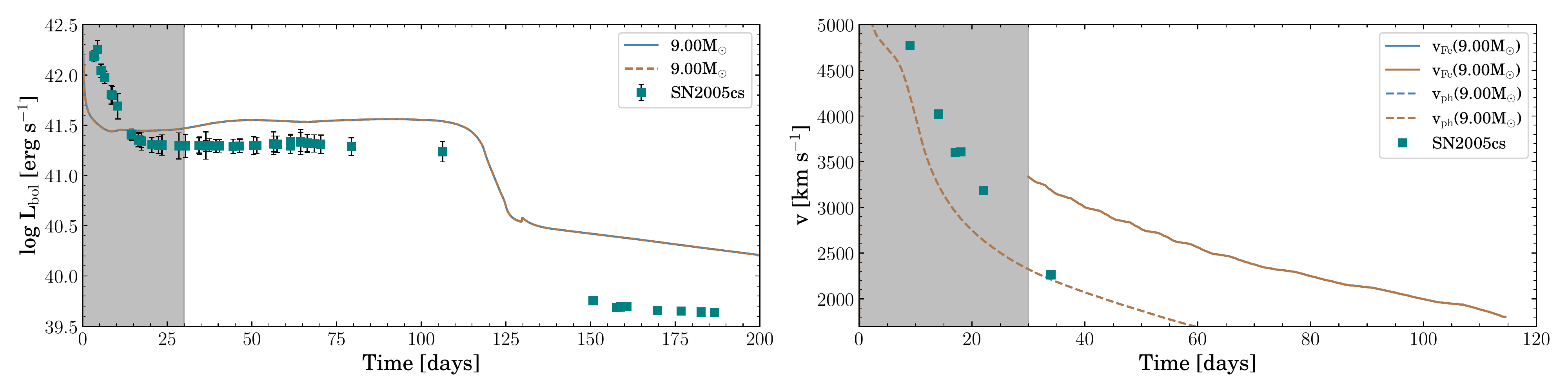} \hfill
  \includegraphics[width = 1.\textwidth]{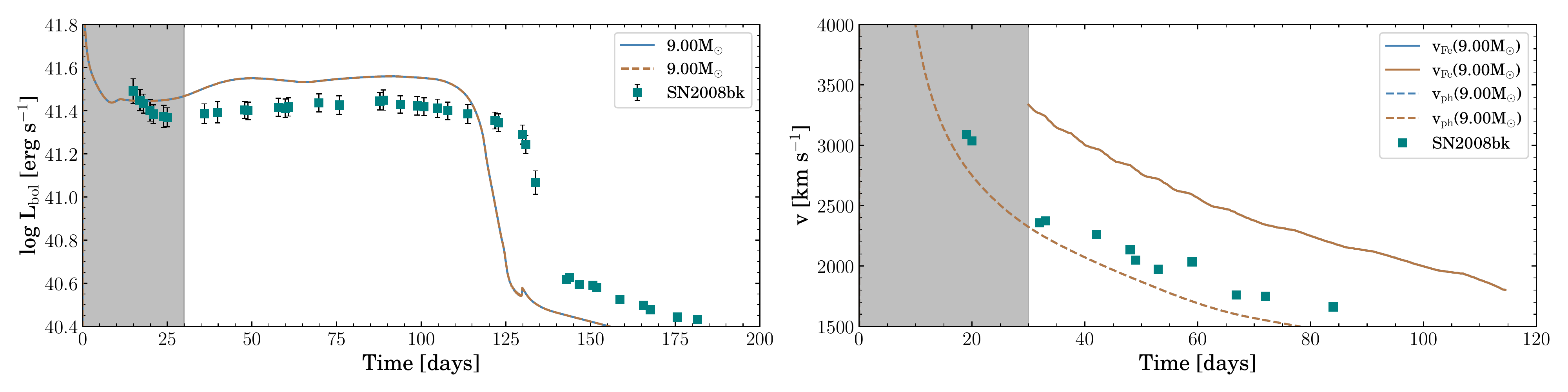} \hfill
  \caption{Left: Comparison between STIR + SNEC light curves (blue lines) and observations (squares).
  Right: Comparison between STIR + SNEC velocity evolution (lines) and Fe II $\lambda$5169$\AA$ line velocity observations (squares).
  Solid lines show approximate Fe II $\lambda$5169$\AA$ calculated in post-processing and dashed lines show the proper photospheric velocity.
  In both plots, blue lines show best fit STIR + SNEC models and gold lines show light curves for ZAMS masses obtained from pre-explosion imaging \citep{smartt:2015, davies:2018, eldridge:2019}.
  The gray shaded region shows the first 30 days that we omit from fitting.
  \textit{From top to bottom}: SN 2004A, SN 2004et, SN 2005cs, and SN 2008bk.}
  \label{fig:lc_fits1}
\end{figure*}

\begin{figure*}
  \centering
  \includegraphics[width = 1.0\textwidth]{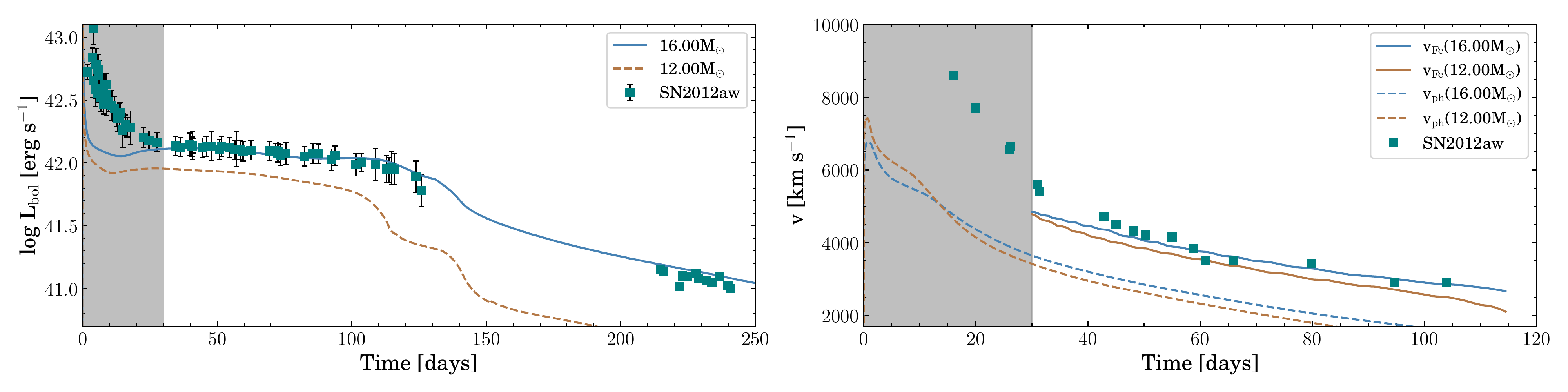} \hfill
  \includegraphics[width = 1.0\textwidth]{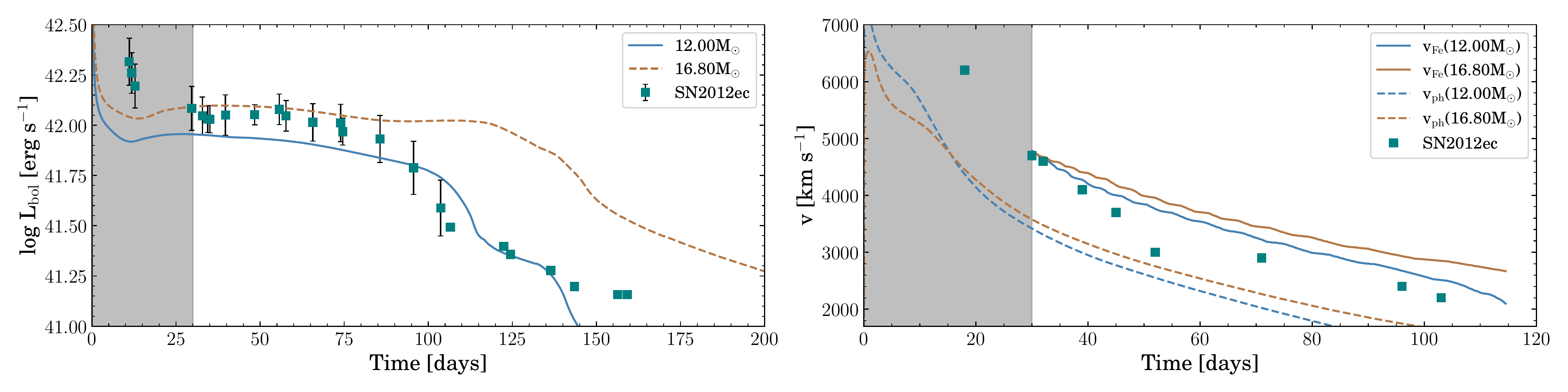} \hfill
  \includegraphics[width = 1.0\textwidth]{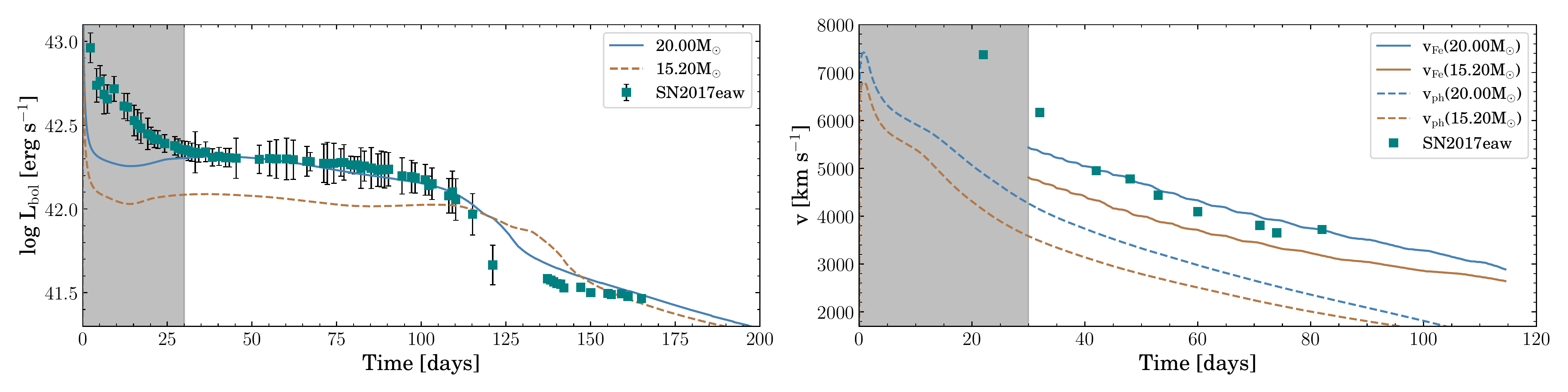} \hfill
  \caption{Same as Figure~\ref{fig:lc_fits1} but for 
  \textit{(from top to bottom)}: SN 2012aw, SN 2012ec, and SN 2017eaw}
  \label{fig:lc_fits2}
\end{figure*}

Figures~\ref{fig:lc_fits1} and \ref{fig:lc_fits2} show observed bolometric light curves (left) and velocity evolution (right) for (top to bottom) SN 2004A, SN 2004et, SN 2005cs, SN 2008bk, SN 2012aw, SN 2012ec, and SN 2017eaw.
Dark blue lines show bolometric luminosity and velocity evolution for best fit progenitors from our sample using the STIR + SNEC model using the fitting described in Section~\ref{sec:fitting}.
For the velocity evolution, dashed lines show approximate Fe II $\lambda$5169$\AA$ line velocities estimated through the methods described in Section~\ref{sec:observations} and solid lines show the proper $\tau = 2/3$ photospheric velocity.
Gold lines are for ZAMS mass models corresponding to estimates from pre-explosion imaging. 
We use the ZAMS mass estimates from \citet{davies:2018} for SN 2004A, SN 2004et, SN 2008bk, SN 2012aw, and SN 2012ec. 
Properties of these SNe are discussed in detail in \citet{martinez:2019} and \citet{martinez:2020}.
For SN 2005cs, \citet{davies:2018} estimated an initial mass of about 7 \Msun -- well below the minimum mass we consider to produce a CCSN -- so we use the estimate from \citet{smartt:2015}.
Finally, we use the mass estimate for SN 2017eaw from \citet{eldridge:2019}.
In all cases we use the optimal value of the initial mass when possible, or the closest value within the reported range that was both on our mass grid and produced an explosion.

We determine the best fit progenitor by minimizing the total relative error of both luminosity and velocity across the entire light curve after day 30 as discussed in Section~\ref{sec:fitting}. 
We also tried minimizing $\chi^2$, as was done in \citet{morozova:2018b}, but found unsatisfactory performance compared to our method (see Appendix~\ref{app:app2} for an example using SN2017eaw).
We did not consider the errors associated with the observations in our fitting. 
The inverse variance weighting typically used in $\chi^2$ minimization gave stronger significance to the radioactive tail, as this region has much smaller error compared to the plateau.
The result was the selection of models that fit the tail nicely, but fit the plateau very badly.
We do not consider data before 30 days post shock breakout, as very early time bolometric luminosities may be heavily influenced by interactions with circumstellar material (CSM) for some SNe \citep{morozova:2018b} and we have not included CSM effects in this work.

We do not expect to find close fits for all observed CCSNe.
In this work, we have progenitors that cover a wide range of ZAMS masses with explosions driven by turbulence-aided neutrino radiation hydrodynamics simulations, but are limited in scope in other regards, such as rotation, metallicity, $^{56}$Ni mass and distribution, and possible effects of binarity.
Moreover, we do not have models with masses lower then 9\Msun, which may contribute to CCSNe.
For example, SN 2008bk is very underluminous with low expansion velocities as is very likely a lower mass progenitor than we have in our set \citep{mattila:2008, van-dyk:2012b, lisakov:2017, lisakov:2018, martinez:2019, oneil:2020}.
With these limitations in mind, we still find good fits for two observed CCSNe, notably 16.0 \Msun for SN 2012aw and 20.0 \Msun for SN 2017eaw.
Our best fit progenitors tend to have larger ZAMS masses than those estimated from pre-explosion imaging, for example by about 5\Msun for SN2017eaw.
This difference of about 5\Msun is not uncommon -- \citet{goldberg:2020b}, for example, find a possible ZAMS mass for SN2017eaw of 10.2\Msun -- also about 5\Msun from the value obtained from pre-explosion imaging.
We have presented light curves for which we do not find particularly good fits for the sake of completeness and to show the strengths and weakness of the current progenitor set.
As previously mentioned, there is no reason for this progenitor set to perfectly fit any specific light curve.

The differences highlighted in Figures~\ref{fig:lc_fits1}~and~\ref{fig:lc_fits2} show the inherent degeneracy involved in extracting CCSNe progenitor properties.
As shown in \citet{goldberg:2019, dessart:2019}, there are familes of progenitor properties that can lead to a given light curve.
This further highlights that light curve fitting is extremely degenerate -- not only in the ways explored in previous works, but also in the method used to drive the explosion.
Thus, we do not claim that these progenitors necessarily reflect the true progenitors, they simply match the observations given a set of neutrino-driven explosions.
It has become clear that more work in needed to infer progenitor properties.
Matching an observed SN is a necessary, but not sufficient, condition for inferring progenitor and explosion properties.

Finally, we summarize our best fitting models for those light curves for which we see good agreement (SN2012aw and SN2017eaw) in Table~\ref{table:summary} alongside various other sources.

\begin{table*}
  \centering
  \topcaption{Best-fit ZAMS mass and explosion energy for SN2012aw and SN2017eaw for our work and others in the literature.
   For all works we present the best fit model reported with the exception of 5 \citep{goldberg:2020b}, where we list all presented matches.\\
   $^{\dagger}$ The authors only present ejected mass, so we present that as a lower bound for the ZAMS mass.} 
  \label{table:summary}
  \begin{tabular}{l l c c c c c c }
    \toprule
     SN & Quantity & This work & 1 & 2 & 3 & 4 & 5 \\
     \hline
     \multirow{2}{*}{2012aw} & M$_{\mathrm{ZAMS}}$ [\Msun] & 16.0 & 23.0 & 14.35 & 20.0 & $>$19.6$^{\dagger}$ & -- \\ 
             & $E_{\mathrm{expl}}$ [10$^{51}$ erg] & 0.66 & 1.40 & 0.90 & 0.52 & 1.5 & -- \\
     \multirow{2}{*}{2017eaw} & M$_{\mathrm{ZAMS}}$ [\Msun] & 21.9 & -- & 15.47 & -- & -- & (10.2, 12.7, 17.2, 11.9, 15.7, 19.0) \\ 
             & $E_{\mathrm{expl}}$ [10$^{51}$ erg] & 1.09 & -- & 1.29 & -- & -- & (0.65, 0.84, 1.30, 0.90, 1.10, 1.50) \\ 
    \midrule

    \bottomrule

  \end{tabular}
  \newline
  \textbf{References: (1) \citet{martinez:2019}; (2) \citet{martinez:2020}; (3) \citet{morozova:2018b}; (4) \citet{pumo:2017}; (5) \citet{goldberg:2020b}.}
\end{table*}

\subsection{Correlations}
\label{sec:correlations_results}

\begin{figure}
  \centering
  \begin{tabular}{c c}
  \includegraphics[width = 0.45\textwidth]{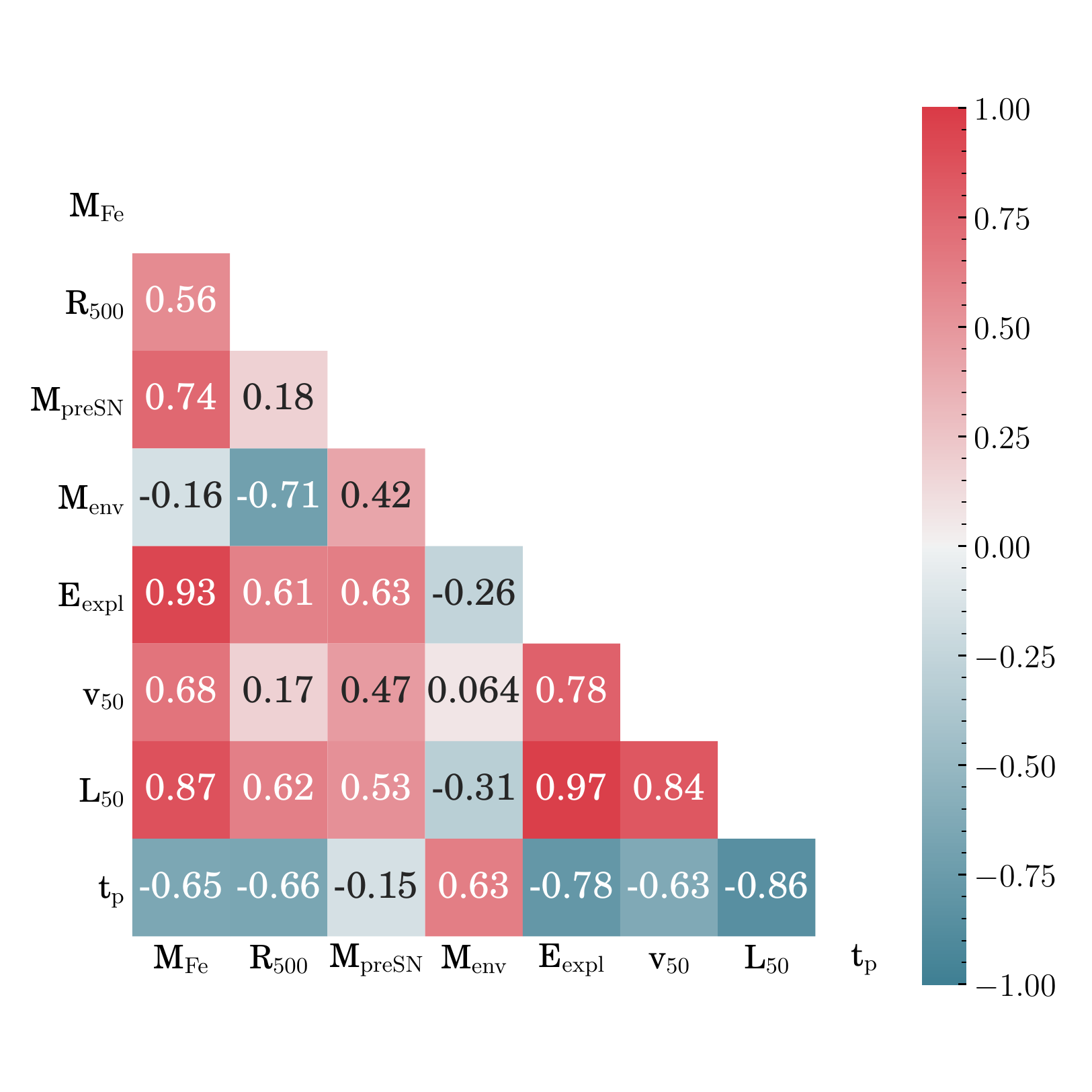}
  \end{tabular} 
  \caption{Correlation matrices for observable quantities and properties of the progenitors for STIR + SNEC. 
  Here we consider the following quantities: iron core mass ($M_{\mathrm{Fe}}$), progenitor radius ($R_{500}$), explosion energy ($E_{\mathrm{expl}}$), 
  ejecta velocity at day 50 ($v_{50}$) as determined from the Fe II ($5169\AA$) line, log of the plateau luminosity 
  at day 50 ($L_{50}$), and plateau duration ($t_{\mathrm{p}}$). The lower left half of the matrix shows the Spearman rank correlation coefficient for each pair of quantities.
  }
  \label{fig:corr_primary} 
\end{figure}

In this section, we address the primary goal of this study, which is to connect light curve properties to progenitor properties using a statistically significant sample of simulations.
Figure~\ref{fig:corr_primary} shows the Spearman's correlation matrix for the observable quantities and progenitor properties that we consider for the STIR + SNEC models. 
Our goal is to assess direct correlations between individual quantities, and for this reason we do not consider correlations with ZAMS mass because it does not correlate with any single quantity.
In many cases, we are simply recovering well-known correlations, which provide a sanity check on our methods.
For example, relationships between ejecta velocity and luminosity have been used in SNe IIP supernova cosmology \citep{hamuy:2005, nugent:2006, poznanski:2009}. 
Relationships between photometric and spectroscopic observables, $L_{50}$, $v_{50}$, and $t_{p}$, and properties of the progenitor, such as $R_{500}$ (the pre-supernova progenitor radius in units of 500$R_{\odot}$) in addition to the explosion energy are used in scaling relationships, such as those in \citet{popov:1993, kasen:2009a,sukhbold:2016,goldberg:2019}.

We first consider some typical observables of SNe IIP light curves -- the plateau luminosity ($L_{50}$), plateau duration ($t_p$), and ejecta velocity measured through the Fe 5169$\AA$ line during the plateau phase ($v_{50}$).
These observables correlate with each other and are expected to correlate with properties of the progenitors, such as the presupernova radius ($R_{500}$) and envelope mass ($M_{\mathrm{env}}$).
We observe significant correlations between $t_p$, $L_{50}$, and $R_{500}$.
Correlations with $R_{500}$ tend to be non-monotonic (see, e.g., Figure~\ref{fig:progenitors}), which is why they tend to
have weaker values of the correlation coefficient.
There is a moderate correlation between the $L_{50}$ and $v_{50}$ and the presupernova mass ($M_{\mathrm{preSN}}$).

The explosion energy ($E_{\mathrm{expl}}$, see Section~\ref{sec:FLASH}) is expected to correlate with both progenitor properties and observable properties.
Correlations between $E_{\mathrm{expl}}$ and observable properties are monotonic relationships (i.e., always increasing or always decreasing, but not necessarily linear), for example with a correlation coefficient of 0.97 for $L_{50}$ -- $E_{\mathrm{expl}}$.
This is because in the self-consistent STIR + SNEC models, the explosion energies are the total positive energies of unbound material as liberated by neutrino heating and is thus correlated with properties of the core (and thus, the rest of the progenitor properties through stellar evolution) of the progenitor.

Finally, we turn our attention to connections between properties of the core of the progenitor and observable quantities.
Motivated by connections between explosion energy and the compact remnant, we explore correlations with the iron core mass ($M_{\mathrm{Fe}}$).
Progenitors with more massive iron cores tend to liberate more gravitational binding energy, have higher neutrino luminosities, and ultimately are associated with more energetic explosions for progenitors that successfully explode.
The origins of this correlation can be seen in the bottom panel of Figure~\ref{fig:e_expls} through the connection between iron core mass and explosion energy.
This correlation, therefore, once again highlights the need for realistic physics in explosion models even in 1D.
Equipped with this correlation, and the previously mentioned relationships between explosion energy and observables, one might expect some imprint of the iron core mass on the observables.
Indeed, for the STIR + SNEC models we observe a very strong, linear relationship between iron core mass and plateau luminosity.
We note that the compactness parameter $\xi_{2.5}$ \citep{oconnor:2011} produces a stronger correlation.
This, however, is of little practical use, as the 9-12\Msun progenitors have nearly zero values of the compactness parameter ($\leq$ 0.02), breaking the trend for the most common progenitors, and the iron core mass is a more physical quantity (i.e., does not depend on the exact choice of mass coordinate for the measurement).
The compactness parameter and iron core mass are very tightly correlated and both provide a measure of the gravitational binding energy available in the explosion.

\begin{figure}
  \centering
  \includegraphics[width = 0.45\textwidth]{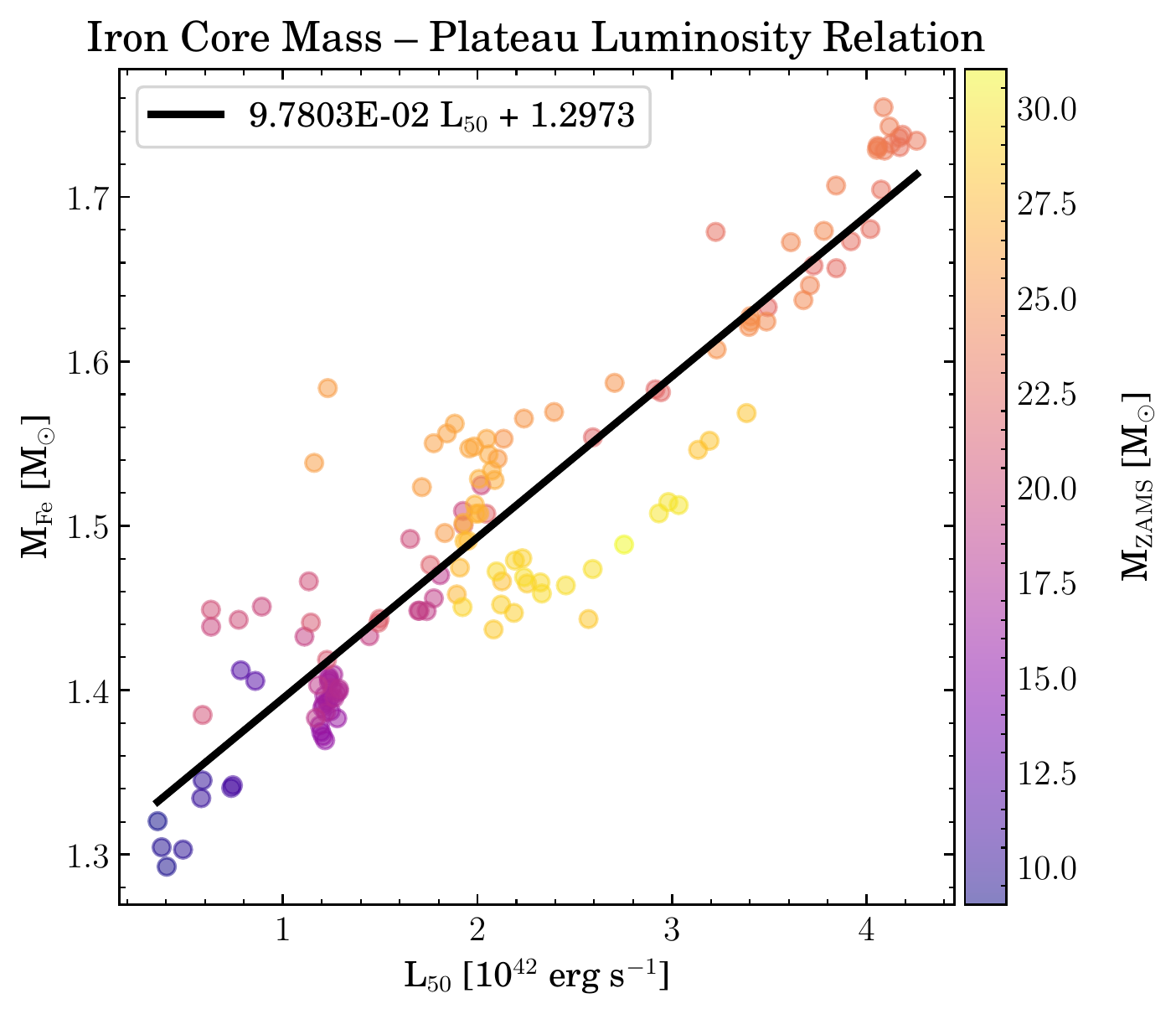}
  \caption{Iron core mass $M_{\mathrm{Fe}}$ versus plateau luminosity at day 50 $L_{50}$.
  }
  \label{fig:MFe_L}
\end{figure}

A relationship between iron core mass and supernova observables helps constrain stellar evolution models and characterize the diversity of supernova light curves.
Figure~\ref{fig:MFe_L} shows iron core mass versus plateau luminosity at day 50.  
Higher luminosity events tend to originate from progenitors with more massive iron cores.
Ultimately, more massive stellar cores collapse to form more massive proto-neutron stars, liberating more gravitational binding energy in the process and resulting in higher neutrino luminosities emanating from the PNS surface.
All of this results in a more energetic explosion and a brighter supernova. 
In Table~\ref{table:fits} we report the fits coefficient for the $M_{\mathrm{Fe}}$-$L_{50}$ relationship and the associated variances and covariances for a linear fit of the form $y = ax + b$.

This correlation, though simple, has a profound implication that we can constrain core structure from optical photometry alone. 
While not necessarily providing a \textit{precise} measure of the iron core mass for individual events due to observational error and uncertainties on the fit parameters from scatter, which we quantify below, it provides a method for comparing the cores of \textit{virtually all} SNe IIP simultaneously. 
Furthermore, these parameter estimates can  be used to constrain stellar evolution models for CCSN progenitors.
We find a similar, although slightly weaker, correlation between the ejecta velocity at day 50, $v_{50}$, as well, but most LSST sources won't have a spectroscopic follow-up so this is of limited use.

For the case of \citet{sukhbold:2016}, fewer massive iron cores produced explosions, and the explosions had a tendency to be brighter.
Using their data we find slope and intercept parameters of 0.033 and 1.344, respectively.

For any relationship of this type to be useful, error must be taken carefully into account. 
The optimal fit parameters were obtained with a least squares method.
However, it is known that the covariances provided by least squares methods are not appropriate for a wide range of problems, including those with a non-Gaussian intrinsic scatter among other criteria \citep[see, e.g.,][and references therein]{clauset:2009}.
For this reason, we resort to a bootstrapping method \citep{efron:1979} to obtain the errors on the fit parameters. 
This method has the advantage of making no assumptions about the underlying distribution of the data.
Instead, bootstrapping operates by resampling the data $M$ times with replacement.
For each resampling, a new fit is made and those fit parameters stored. 
Then, estimates of the variance and covariance of parameters $u$ and $v$ are given by

\begin{equation}
    \sigma_{u}^2 = \frac{1}{M} \sum_{j=1}^{M} \left( u_j - u \right)^{2} \\
\end{equation}
\begin{equation}
    \sigma_{uv} = \frac{1}{M} \sum_{j=1}^{M} \left( u_j - u \right)\left( v_j - v \right)
\end{equation}

\noindent where $u$ and $v$ are the optimal fit parameters and each of $u_j$, $v_j$ are the fit parameters for each of the $M$ resamples.
These error estimates tend to be, for this application, somewhat smaller than parameter errors obtained through a simple least squares method.
The full set of fit parameters, variances, covariances, and adjust coefficient of determination are supplied in Table~\ref{table:fits}.
We note that the fit presented uses the non-log plateau luminosity as its independent variable, as opposed to the log luminosity presented in other parts of the paper.
Then, given errors on the fit parameters it is straightforward to compute the error on an iron core mass estimate. 
For a linear fit, we propagate the combined observational -- fit parameter uncertainty in the following way:

\begin{equation}
\sigma_{\mathrm{M_{Fe}}}^2 = \sigma_{a}^2 L_{50}^2 + \sigma_{L_{50}}^2 a^2 + \sigma_{b}^2 + \sigma_{\mathrm{res}}^2 + 2 L_{50} \sigma_{ab},
\end{equation}
\noindent where $L_{50}$ is the luminosity at day 50 in erg s$^{-1}$ and where we have included explicitly the covariance of the fit parameters $a$ and $b$.
In order to further account for intrinsic scatter in the relationship, we have included $\sigma_{\mathrm{res}}$ which is the 67$\%$ percentile on the residual distribution $r_{i} = \lvert M_\mathrm{{Fe}} - \hat{M}_\mathrm{{Fe}}\rvert$, where $\hat{M}_\mathrm{{Fe}}$ is computed from the fit.

\begin{table}
  \centering
  \topcaption{Linear fit paramaters for iron core mass ($M_{\mathrm{Fe}}$) to plateau luminosity ($L_{50}$){in units of 10$^{42}$ erg s$^{-1}$.
      The first two rows shows the optimal fit parameters.
      The next two rows shows the error on each parameter.
      The next row shows the covariance between the parameters and the residual error accounting for intrinsic scatter.
      The final row shows the adjusted coefficient of determination $\bar{R}^{2}$ for the fit.}} 
  \label{table:fits}
  \begin{tabular}{@{}c c@{} }
    \toprule
     & \multicolumn{1}{l}{$M_{\mathrm{Fe}}$ = $a L_{50} + b$}  \\
    \midrule
      $a$ & 0.0978 \\
      $b$ & 1.29  \\
    \midrule
      $\sigma_a$ & 3.17$\times 10^{-3}$  \\
      $\sigma_b$ & 8.31$\times 10^{-3}$  \\
    \midrule
      $\sigma_{ab}$ & -2.33$\times 10^{-5}$ \\
      $\sigma_{\mathrm{res}}$ & 3.79$\times 10^{-2}$ \\
    \midrule
      $\bar{R}^{2}$ & 0.85 \\
    \bottomrule

  \end{tabular}
\end{table}

As an example, we estimate iron core masses for six well-observed SNe, shown in Table~\ref{table:mfe}. 
Data for SN1999em, SN2003hl, and SN2007od are taken from \citet{gutierrez:2017, gutierrez:2017a}.
Data for SN2004et, SN2012aw, and SN2017eaw are taken from \citet{martinez:2020}.

\begin{table}
  \centering
  \topcaption{Estimated iron core masses ($M_{\mathrm{Fe}}$) and uncertainties ($\sigma_{\mathrm{M_{Fe}}}$) for a sample of well-observed supernovae.} 
  \label{table:mfe}
  \begin{tabular}{l c c }
    \toprule
     SN & $M_{\mathrm{Fe}}$ [\Msun] & $\sigma_{\mathrm{M_{Fe}}}$ [\Msun] \\
    \midrule
      1999em  & 1.42 & 0.041 \\
      2003hl  & 1.34 & 0.039 \\
      2004et  & 1.48 & 0.039 \\
      2007od  & 1.50 & 0.040 \\
      2012aw  & 1.43 & 0.039 \\
      2017eaw & 1.49 & 0.040 \\
    \bottomrule

  \end{tabular}
\end{table}

\section{Discussion and Conclusions}
\label{sec:Conclusions}

We present synthetic bolometric light curves for 136 solar metallicity, non rotating CCSNe progenitors and consider statistical relationships for those with ZAMS masses ranging from 9\Msun to 31\Msun. 
These light curves are calculated with SNEC using the CCSNe simulated in \citet{couch:2020} as the initial condition.
This allows for light curves obtained without a user-set explosion energy.
Our $^{56}$Ni yields were fit from \citet{sukhbold:2016} who exploded the same progenitors with an expansive reaction network coupled to the evolution.
This is sufficient for the current work, and future work with \texttt{FLASH} will include detailed nucleosynthesis calculations.
These light curves, as well as the SNEC initial profiles and necessary parameters, are provided online\footnote{\url{https://doi.org/10.5281/zenodo.6631964}}.
We also include the necessary binding energy of our progenitors to correct STIR's explosion energy to produce identical results with a thermal bomb explosion.
In the online resources, we furthermore provide the light curves for the $M_\mathrm{ZAMS}>31\Msun$ models that successfully explode.
For progenitors that explode with STIR, we follow the explosions in SNEC to produce bolometric light curves, forming a large, statistically significant set of CCSN light curves followed from high-fidelity explosions allowing us to address relationships between progenitor properties and properties of the explosion in a statistical way.
We consider the full shape of these light curves, but also reduce them to characteristic quantities such as 
the plateau luminosity, plateau duration, and ejecta velocity.

Next, we show that global trends in light curve properties -- such as plateau duration and plateau luminosity -- depend sensitively on the explosion model and require explosion energies set by robust physics.
To demonstrate this, we compute bolometric light curves for the same set of progenitors using two different thermal bomb models with SNEC. 
The distribution of explosion energies plays a leading role in setting the distribution of observables across a large sample of progenitors.
Thus, the ability to identify global trends in light curve properties and extract progenitor features from them depends sensitively on the determination of explosion energy, underscoring the need for explosions driven with high-fidelity multi-physics models.

We present a simple best-fit procedure to individual, observed CCSN light curves \citep{martinez:2020}.
The usual procedure for estimating progenitor properties of observed CCSNe is to construct a large grid of ``hydrodynamical models'' -- usually in ZAMS mass, explosion energy, and perhaps $^{56}$Ni mass and distribution -- and find a best fit model.
This approach results in known degeneracies, for example, as shown by \citet{goldberg:2019, dessart:2019} wherein 
there are certain families of progenitor and explosion parameters (such as ejecta mass, explosion energy, and ejecta velocity) that produce a given light curve, though pre-explosion radius measurements may help to resolve this degeneracy \citep{goldberg:2020b, kozyreva:2020a}.
Our approach differs in that we do not control the explosion properties, instead following a dense set of various ZAMS mass progenitors from neutrino driven explosions.
While this does not solve the light curve degeneracy problem, it could reduce the size of the family of explosion properties for a given light curve, as some combinations of explosion energy and stellar mass are not realizable.
Although the explosions are not calibrated to observed data we still find great agreement both when comparing to large samples of events and for some individual cases. 
Intriguingly, we find best-fit ZAMS masses that are greater by as much as $\approx$7\Msun than those estimated from pre-explosion imaging in tandem with stellar evolution modeling.
The fact that hydrodynamic models have tended to find ZAMS masses in agreement with pre-explosion imaging estimates for these CCSNe \citep{morozova:2018b,martinez:2019, martinez:2020} may indicate the danger of exploring too large a parameter space instead of knowing which regions are physically realizable, though we note that some hydrodynamic models have also found noticeably higher masses in better agreement with our conclusions \citep[e.g.,][]{utrobin:2008,utrobin:2009}.
Ultimately, the set of solutions for matching a given observed light curve is degenerate, with many progenitors being capable of producing a given light curve.

Despite the progenitors and explosions in this study not being crafted to reproduce specific events, we find good qualitative agreement with SN2012aw and SN2017eaw.
Notably, the luminosity evolution of SN 2012aw is fit by our 16.0\Msun progenitor remarkably well.
The best fit progenitors for the observed light curves in this study are not necessarily the progenitors that these explosions originated from -- they simply reproduce the observables.
We have demonstrated that beyond the now understood light curve degeneracies, there are additional degeneracies inherited from the choice of explosion model.
This result is complementary to the recent findings by \citet{farrell:2020} where they showed that a star's final temperature and luminosity cannot be reliably traced back to the star's ZAMS mass -- that very different mass stars may end up at the same temperature and luminosity.
These results together show that much more work is needed before a SN IIP progenitor's ZAMS mass can be reliably determined -- the path from stellar birth to death is not a one-to-one function.

The light curves here present avenues for future work to explore the discussion surrounding explosion energy. 
There is tension between explosion energies realized in 3D CCSN simulations and energies inferred from fitting hydrodynamical models to observations.
The energies from these two methods differ, with those inferred from hydrodynamical modeling being significantly larger \citep[see][which discusses this tension in detail]{murphy:2019}.
On one hand, 3D simulations of very massive progenitors have often simply not asymptoted to their final values within the simulated time.
There is also still physics left to include, such as the recently demonstrated affects of magnetic fields on neutrino-matter interactions \citep{kuroda:2020} and improved neutrino pair-production rates \citep{betranhandy:2020} on the explosion mechanism, neutrino mixing, among other affects, all of which will likely play a role in setting the final energy.
On the other hand, solutions using thermal bomb models have been shown to be degenerate, and these studies access a very large area of this degenerate parameter space and may not necessarily find physically realizable solutions.
The methods described here could illuminate or even weaken the tension between these energies by limiting the parameter space spanned by hydrodynamical modeling studies and by using physically-motivated explosions.

The final aim of this study is to leverage the large number of light curves to perform a statistical investigation of relationships between progenitor and explosion properties.
Focusing our investigation to SNe II, 136 light curves, a number of correlations between the light curves and their progenitors are found.
We find a robust relationship between the iron core mass of the progenitor and the luminosity on the plateau of the SNe.
This relationship allows one to, for the first time, constrain properties of the stellar interior from photometry alone.
We provide an analytic approximation to the observed correlation, including error, for future use with large survey data such as LSST.

Recently, \citet{curtis:2020} presented synthetic light curves and spectra from a sample of 62 CCSNe with the 1D PUSH model \citep{ebinger:2017} and SNEC to obtain the light curves.
Our results complement one another in several ways -- notably, the size and composition of our samples differ.
Our sample contains 148 light curves -- 136 of which are analyzed in this work -- from the same metallicity, compared to their 62 light curves from three different metallicity populations ranging from zero to solar.
This allows us to more robustly survey global explosion properties of progenitors from similar origin within the nearby universe.
These studies, together, survey a vast range of progenitor properties.
The CCSNe simulations in our work are perfmormed with \texttt{FLASH} using the STIR model.
Notably, STIR requires no tuning to observations, eliminating the potential for biases when simulating progenitors different than the one used for tuning.
Importantly, the results from STIR are consistent with 3D simulations. 
The explosion energies, explodability, and the shape of each as a function of ZAMS mass differ non-trivially for STIR and PUSH \citep[see][]{couch:2020, ebinger:2018} and this could impact global trends in explosion properties.
On the other hand, \citet{curtis:2020} obtained their $^{56}$Ni distributions using a nuclear reaction network in conjunction with their CCSN simulations.
As aforementioned, we estimated $^{56}$Ni mass from the explosion energy, informed by KEPLER yields.
\citet{curtis:2020} also have a larger diversity of supernova types through their inclusion of sub-solar and zero metallicity progenitors.
To keep the scope of the current work contained, we have not produced synthetic spectra for these explosions, whereas \citet{curtis:2020} calculated spectra for their supernovae. 

Similarly, \citet{sukhbold:2016} present a sample of synthetic light curves of the same statistical size and originating from the same progenitors using a different parametrized, neutrino-driven explosion mechanism.
Using these simulations they present scaling relations to determine explosion and progenitor properties from observables.
The outcomes of these simulations -- both the explosions and resulting light curves -- differ from STIR and this work, having a tendency to be brighter than those produced in this work.
It would be interesting, for future work, to investigate the affect of these differences in explosion mechanism when applied to populations of observed CCSNe and implications for inferred properties such as explosion energy.

This work is part of a larger context to \textit{understand} and \textit{predict} full multi-messenger signals from realistic CCSNe.
Understanding how variations in progenitors properties tie into variations of different observables will ultimately help to constrain real populations.
This work, in tandem with the work of \citet{couch:2020} and \citet{warren:2020}, gives us explosion fates, energies, neutron star mass distributions, neutrino signals, approximate GW signals, and now EM signals for a massive suite of neutrino driven CCSNe.
It is only through advanced methods -- studying in detail all messengers from first principles simulations -- used in tandem with growing observational data that we can truly understand these phenomena.

\software{
\texttt{FLASH} \citep{fryxell:2000,dubey:2009}, 
SNEC \citep{morozova:2015, morozova:2016c}, 
Pandas \citep{mckinney:2010}, 
NumPy \citep{numpy}, 
SciPy \citep{jones:2001} }

\acknowledgments{
We thank E.~H.~Miso for many constructive discussions.
We also thank the anonymous reviewers for their constructive feedback.
BLB is supported by the National Science Foundation Graduate Research Fellowship Program under grant number DGE-1848739.
C.E.H. is grateful for support from NSF through AST-1751874 and AST-1907790, and from the Packard Foundation.
MLW was supported by an NSF Astronomy and Astrophysics Postdoctoral Fellowship under award AST-180184.
EOC is supported by the Swedish Research Council (Project No. 2018-04575 and 2020-00452)
SMC is supported by the U.S. Department of Energy, Office of Science, Office of Nuclear Physics, Early Career Research Program under Award Number DE-SC0015904. This material is based upon work       supported by the U.S. Department of Energy, Office of Science, Office of Advanced Scientific          Computing Research and Office of Nuclear Physics, Scientific Discovery through Advanced Computing     (SciDAC) program under Award Number DE- SC0017955. 
This research was supported by the Exascale Computing Project (17-SC-20-SC), a collaborative effort of two U.S. Department of Energy organizations (Office of Science and the National Nuclear Security Administration) that are responsible for the planning and preparation of a capable exascale ecosystem, including software, applications, hardware, advanced system engineering, and early testbed platforms, in support of the nation's exascale computing imperative.
This work was supported in part by Michigan State University through computational resources provided by the Institute for Cyber-Enabled Research.
We collectively acknowledge that MSU occupies the ancestral, traditional and contemporary lands of the Anishinaabeg – Three Fires Confederacy of Ojibwe, Odawa and Potawatomi peoples.
}

\bibliography{stirLC}

\appendix

\section{Compositional Dependence}
\label{app:composition}
\input{composition}

\section{$\chi^2$ Light Curve Fitting}
\label{app:app2}
\input{app2}

\end{document}

%% file: composition.tex
\begin{figure*}[b]
    \centering
    \includegraphics[width = 0.925\textwidth]{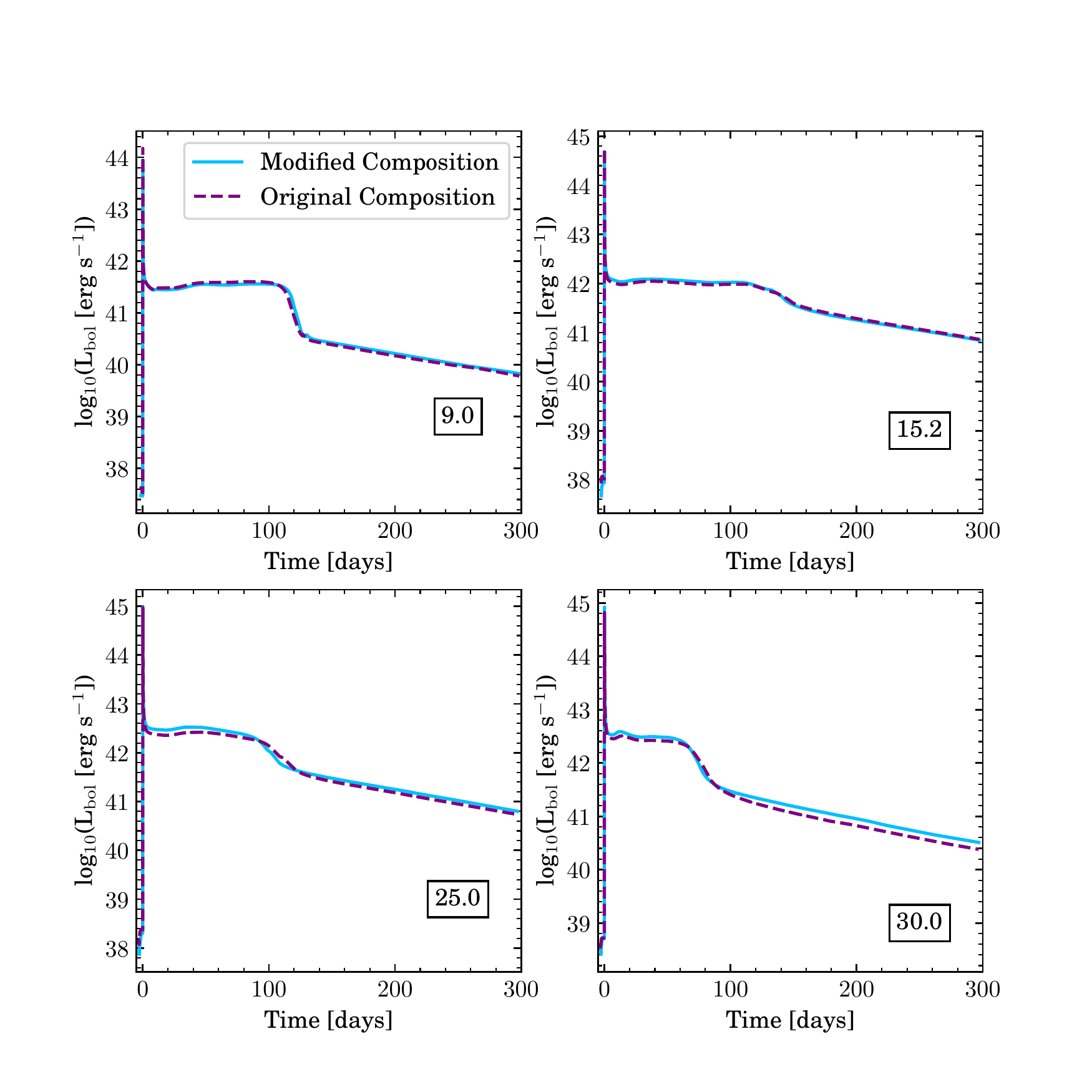}
    \caption{Light curves using a thermal bomb driven explosion with STIR explosion energies using the modified compositional profile (blue) and unaltered profile (orange).
    We show light curves for 9, 15.2, 25, and 30\Msun progenitors.
    }
    \label{fig:comps}
  \end{figure*}

For our light curves, we modified the compositional profile in the FLASH part of the domain to be pure $^{4}$He, as full composition is not currently tracked in the output.
In this appendix, we provide comparisons of select light curves using thermal bombs with FLASH explosion energies for both the modified compositional profile and the original compositional profile.
Figure~\ref{fig:comps} shows light curves with the unaltered (orange) and modified (blue) compositional profiles for 9, 15.2, 25, and 30\Msun progenitors.
For the cases considered here, the difference in luminosity on the plateau is bounded above by 0.1 dex, which has no meaningful affect on the iron core mass estimates and distributions of Section~\ref{sec:correlations_results}.

%% file: app2.tex
\begin{figure*}[h]
    \centering
    \includegraphics[width = 1.\textwidth]{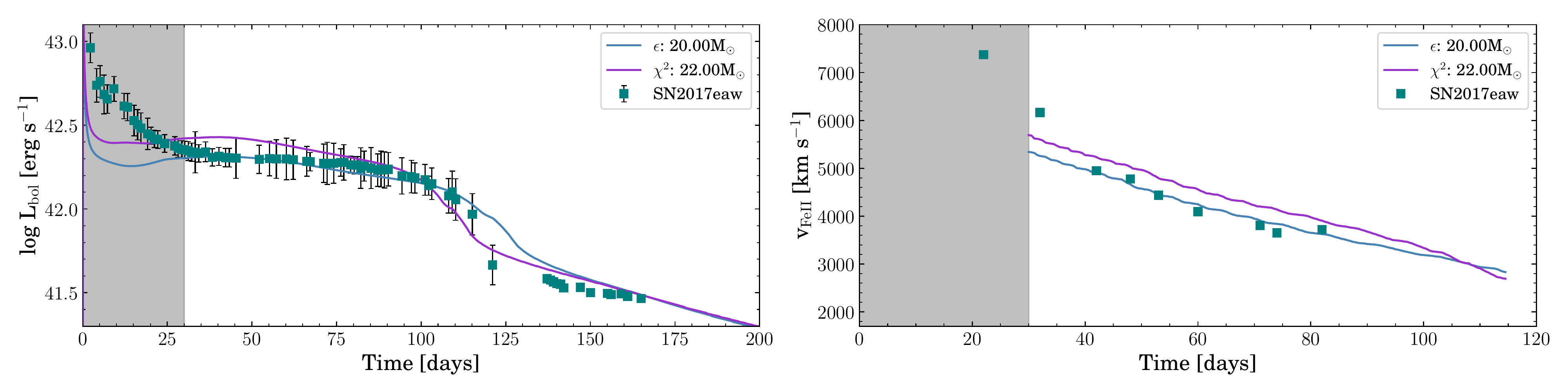}
    \caption{Best fitting light curve for SN1017eaw obtained using a $\chi^2$ metric (purple) and relative error metric (blue).
    }
    \label{fig:chi2}
  \end{figure*}

Here we show the effect of using the $\chi^2$ metric to fit light curves, as opposed to the relative error metric discussed in Section~\ref{sec:fitting}.
We define the chi-square ($\chi^2$) metric for an observable quantity $f\left( t \right)$ as follows

\begin{equation}
  \chi^2(f) = \sum_{t^*=t_1}^{t_N} \frac{\left( f_{t^*} - f_{t^*}^* \right)^2}{\sigma_f^2}
  \label{eq:chi2}
\end{equation}

\noindent where $t^*$ are times coinciding with observations, $f$ are synthetic observables, $f^*$ are measured observational data, and $\sigma_f$ is the uncertainty on the measurement $f^*$ at a time $t^*$.
Here we consider simultaneous fitting of luminosity and velocity data, i.e., minimizing the combined metric $\chi^2(v_{\mathrm{Fe}}) + \chi^2(L_{\mathrm{bol}})$.
Figure~\ref{fig:chi2} shows the best fit model light curve for SN2017eaw using the chi-squared method (purple) and the relative error metric (blue).
The light curve obtained with the chi-square method visibly fits the observations worse than the light curve obtained with the relative error approach, owing to the inverse square error weighting in the chi-square method.
This weighting gives preference to the tail of the light curve where observational errors are reduced.

It is important to note that while chi-square minimization gave less satisfactory results for this study, this is likely sensitive to the details of the data being fit.